\documentclass[aps,prl,twocolumn,floatfix,superscriptaddress]{revtex4-1}

\usepackage{amsmath}  
\usepackage{amsfonts} 
\usepackage{graphicx} 
\usepackage{epstopdf}
\usepackage{gensymb}
\usepackage{physics}
\usepackage{xcolor}
\usepackage{enumerate}
\newcommand{\PT}{\mathcal{PT}}
\newcommand{\K}{\mathcal{K}}
\newcommand{\CP}{\mathcal{CP}}

\newcommand{\vecell}{\skew{2}\hat{\boldsymbol{\ell}}}

\begin{document}
\title{Non-Hermitian band topology and skin modes in active elastic media}

\author{Colin Scheibner}
\affiliation{James Franck Institute, The University of Chicago, Chicago, Illinois 60637, USA}
\affiliation{Department of Physics, The University of Chicago, Chicago, Illinois 60637, USA}
\author{William T.~M.~Irvine}
\affiliation{James Franck Institute, The University of Chicago, Chicago, Illinois 60637, USA}
\affiliation{Department of Physics, The University of Chicago, Chicago, Illinois 60637, USA}
\affiliation{Enrico Fermi Institute, The University of Chicago, Chicago, Illinois, 60637, USA}
\author{Vincenzo Vitelli}
\email{vitelli@uchicago.edu}
\affiliation{James Franck Institute, The University of Chicago, Chicago, Illinois 60637, USA}
\affiliation{Department of Physics, The University of Chicago, Chicago, Illinois 60637, USA}

\begin{abstract}
Solids built out of active components can exhibit non-reciprocal elastic coefficients that give rise to non-Hermitian wave phenomena. 
Here, we investigate non-Hermitian effects present at the boundary of two-dimensional active elastic media obeying two general assumptions: their microscopic forces conserve linear momentum and arise only from static deformations. Using continuum equations, we demonstrate the existence of the
non-Hermitian skin effect in which the boundary hosts an extensive number of localized modes. 
Furthermore, lattice models reveal non-Hermitian topological transitions mediated by exceptional rings driven by the activity level of individual bonds. 
\end{abstract}

\maketitle

The microscopic injection of energy into solid media via active, living, or robotic components fundamentally alters their mechanical waves~\cite{Shmuel2020,Brandenbourger2019,Yoshida2019cl,Rosa2020,Ghatak2019Realization, Zhou2020, Scheibner2020}. As with optics~\cite{LiAlu2019, Sounas2017}, topoelectric circuits~\cite{kotwal2019,helbig2019,Hofmann2020,Yoshida2020el}, and open quantum systems~\cite{Wang2019, Lau2018}, the interplay between activity (gain) and dissipation (loss) can often be captured by non-Hermitian operators~\cite{Bergholtz2019,ashida2020,Hatano1997}. In all these contexts, a central question is what happens at the boundary of the system. Like their Hermitian counterparts, non-Hermitian systems have been shown to exhibit topological invariants that ensure localized boundary modes~\cite{YaoSong2018,Yao2018,Kunst2019,Kunst2018,Torres2019,Shen2018,Li2019,Herviou2019,Kawabata2019PRX,Kawabata2019NatCom,Gong2018,Ghatak2019,Budich2019, Jin2019Topological}. However, in some cases, the familiar bulk-boundary correspondence breaks down for non-Hermtian systems. Such systems exhibit the non-Hermitian skin effect, in which an extensive number of modes are localized to the system's boundary~\cite{Lee2019,Lee2019b,Borgnia2020}. 

Here we examine the non-Hermitian wave phenomena that arise from the elastic properties of a class of active solids. In the continuum, non-Hermiticity enters the linear elasticity of a solid through odd elastic moduli, which are active moduli that violate Maxwell-Betti reciprocity~\cite{Scheibner2020}. We show that such odd elastic moduli when combined with anisotropy can give rise to the non-Hermitian skin effect. This effect implies a dramatic localization of vibrational modes to the system's boundary. Furthermore, we take a microscopic view of elasticity by considering 2D lattices composed of active bonds. These bonds, while active, retain two crucial features of Hookean springs: they conserve linear momentum and depend only on changes in their length. We uncover a non-Hermitian topological transition driven by the level of activity. This transition differs qualitatively from its Hermitian counterpart in that it is mediated by exceptional rings. We interpret such rings in terms of geometric changes in particle trajectories necessary to draw energy from non-potential forces.  

\begin{figure*}[t!]
    \includegraphics[width=\textwidth]{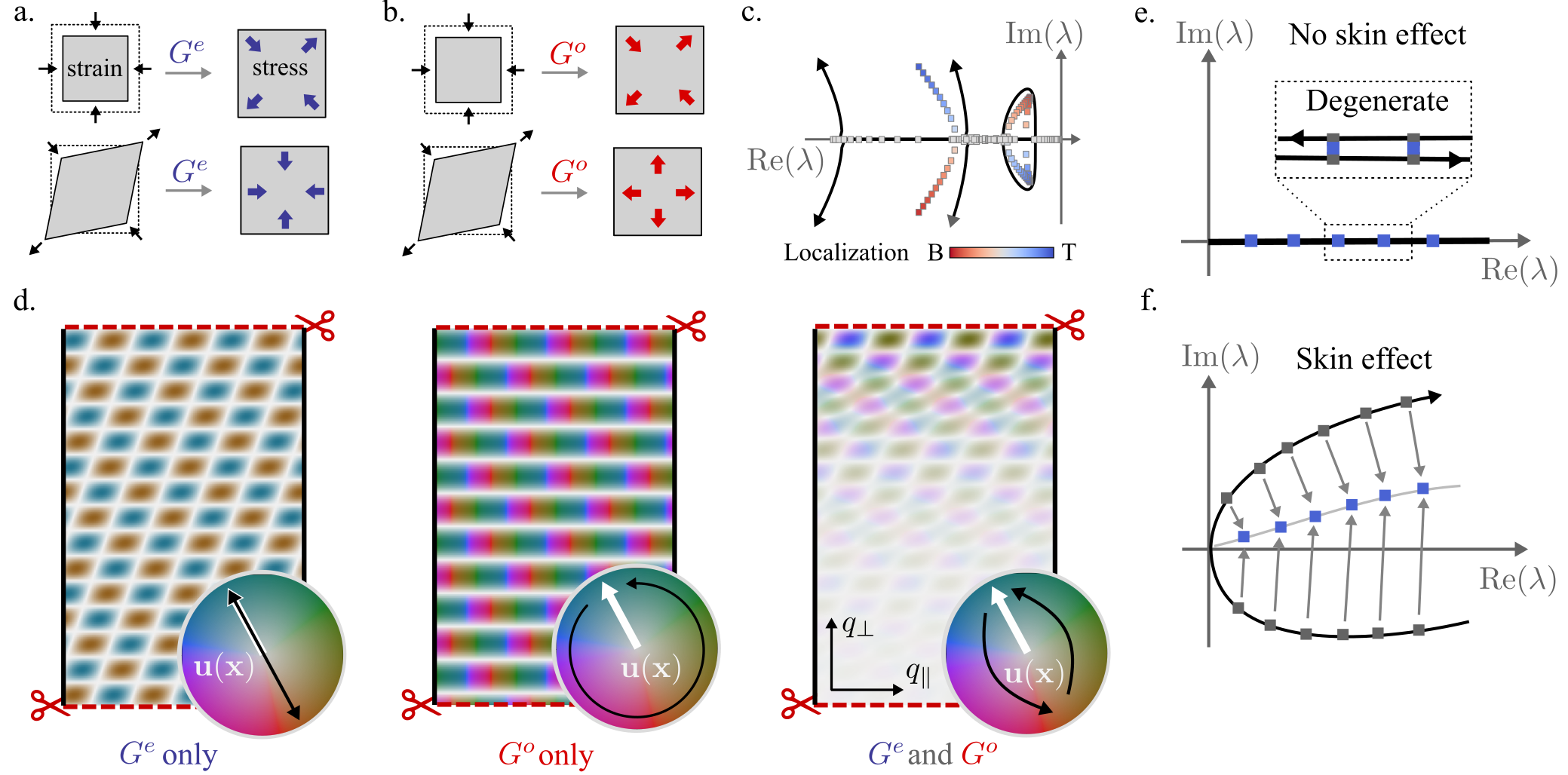}
    \caption{
    {\bf Elastic non-Hermitian skin effect.} {\bf a.}~An anisotropic, passive elastic modulus $G^e$ that couples dilation and shear. {\bf b.}~The corresponding anisotropic, odd elastic counterpart $G^o$. {\bf c.}~The elastic spectrum for $G^o/G^e=1.7$. The black curves (colored squares) correspond to periodic (open) boundary conditions. Blue (red) indicates localization to the top (bottom) boundary.  {\bf d.}~Numerically computed eigenmodes with only $G^e$ present (left), only $G^o$ present (middle), and both $G^o$ and $G^o$ (right). Hue indicates the angle and opacity indicates the magnitude of the displacement field $\vb u(\vb x)$. The horizontal boundaries (red) are open and the vertical boundaries (black) are periodic.
     {\bf e.}~Since the spectrum of a Hermitian system lies on the real line, a generic eigenvalue is at least doubly degenerate. {\bf f.}~For a non-Hermitian system, the spectrum can trace out non-degenerate arcs (black solid). In this case, the spectrum deforms (blue squares) to form degeneracies when a boundary is introduced. 
    }
    \label{fig1}
\end{figure*} 

{\it Non-Hermitian elasticity\textemdash}~We choose as our starting point elasticity theory, the continuum description of solids that captures their ability to resist shape change at large length scales~\cite{Landau7}. Unlike conventional treatments of passive elasticity, we seek to capture at a coarse-grained level the effects induced by {\it non-conservative} internal forces $F_i(\vb x)$ satisfying three assumptions~\cite{Scheibner2020}. First, the forces conserve linear momentum, and therefore can be written as the divergence of a stress $\partial_j \sigma_{ij} (\vb x)$. Second, the forces only depend on the static change in shape, which is captured by gradients of the displacement field $\partial_i u_j (\vb x)$. Finally, we employ the phenomenological assumption that the stresses can be approximated as linearly proportional to the strains: $\sigma_{ij} = C_{ijmn} \partial_m u_n$. The object $C_{ijmn}$ is the elastic tensor, and it encodes the material's response to static deformation.

Following the approach of Ref.~\cite{Scheibner2020}, it is useful to express the elastic tensor as the sum of two pieces:
\begin{align}
    C_{ijmn}= C_{ijmn}^e + C_{ijmn}^o,
\end{align}
where $C_{ijmn}^e=C_{mnij}^e$ is even, or symmetric, under exchange of pairs of the lower indices while $C_{ijmn}^o=-C_{mnij}^o$ is odd, or antisymmetric~\cite{Scheibner2020}. To understand the decomposition, consider the elastic work done (per unit volume) by a patch of material brought through a closed cycle of strain: $w = -\oint \sigma_{ij} \dd u_{ij} = C^o_{ijmn} \oint u_{ij} \dd u_{mn} $.  If the sytem is passive, $w=0$ for any cycle that begins and ends in the same state, and hence $C^o_{ijmn}=0$. However, for an active solid, this requirement does not hold and, consequently, $C^o_{ijmn}$ may be non-zero. We use the term \emph{odd elastic media} to refer to this class of active systems~\cite{Scheibner2020}.

Here, we focus on {\it anisotropic} odd-elastic media and illustate how they generically exhibit the non-Hermitian skin effect. For concreteness, consider a minimal example of anisotropic odd elasticity represented by the following pictorial stress-strain relationship (see S.I. for standard tensor notation):
\begin{align}
    \raisebox{-0.5\height}{\includegraphics{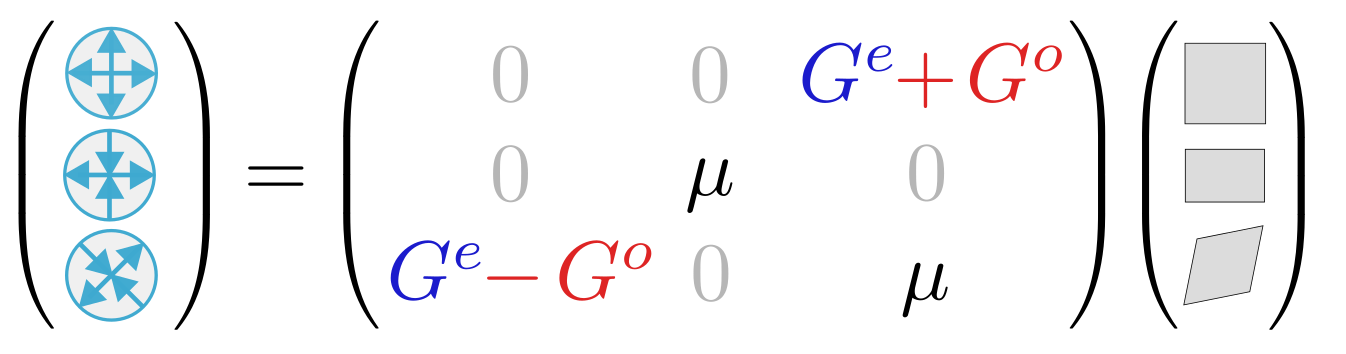}}. \label{mat}
\end{align}
The modulus $G^e$ (contained in $C^e_{ijmn}$) couples dilation 
(\raisebox{-0.2\height}{\includegraphics{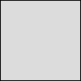}}) 
to shear stress 
(\raisebox{-0.3\height}{\includegraphics{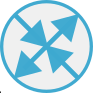}}) 
and shear strain (\raisebox{-0.2\height}{\includegraphics{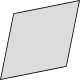}}) 
to pressure 
(\raisebox{-0.3\height}{\includegraphics{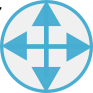}})
symmetrically, see Fig.~\ref{fig1}a. By contrast, the modulus $G^o$ (contained in $C^o_{ijmn}$) provides an antisymmetric coupling (Fig.~\ref{fig1}b). Equation~(\ref{mat}) also includes the standard shear modulus $\mu$ to ensure mechanical stability. In the presence of these moduli, the expression for elastic forces may be written as $\vb F (\vb x)= \hat D \vb u(\vb x)$, where $\hat D$ takes the following form:
\begin{align}
    \hat D = \mqty(\mu \nabla^2 + 2G^e \partial_x \partial_y  & G^e \nabla^2 + G^o [\partial_x^2 - \partial_y^2]  \\ G^e \nabla^2 - G^o [\partial_x^2 - \partial_y^2] &  \mu \nabla^2+ 2G^e \partial_x \partial_y  ), \label{eq:D}
\end{align}
where $\nabla^2 = \partial_x^2 + \partial_y^2$. 
In Eq.~(\ref{eq:D}) we observe that the operator $\hat D$ becomes non-Hermitian when $G^o$ is nonzero, i.e. when the anisotropic moduli display an odd component arising from non-conservative elastic forces. 

We find that the non-Hermiticity has a dramatic effect on the nature of the bulk modes. 
In Fig.~\ref{fig1}c, we show the spectrum of $\hat D$ as a function of wave-number $q_y$ for fixed $q_x$ when both $G^e$ and $G^o$ are present. We find two striking features. First, the open boundary spectrum (square markers) differs dramatically from the spectrum with periodic boundaries (black lines). Secondly, when we examine a typical eigenmode (Fig.~\ref{fig1}d, right) we find that the mode is exponentially localized to the open edge. In Fig.~\ref{fig1}c, we color each mode by the degree of localization to the top (red) and bottom (blue) boundaries. In contrast to typical topological waves or Rayleigh waves in Hermitian systems~\cite{Landau7}, an extensive number of modes are localized to the boundary. 
This extensive localization of bulk modes is an elastic manifestation of the non-Hermitian skin effect.

Yet, when either $G^o=0$ or $G^e=0$, the skin effect disappears (Fig.~\ref{fig1}d left and center, respectively). 
To gain insight into its origins, we consider the notion of a generalized Brillouin zone~\cite{Lee2019,Borgnia2020,Lee2019b}. Let $q_\parallel$ ($q_\perp$) be the wave number parallel (perpendicular) to the boundary. For a finite system, at least two Bloch modes must have the same eigenvalue $\lambda$ and wave number $q_\parallel$ in order to interfere to satisfy a given boundary condition, e.g. $\vb u=0$. If $\hat D$ is Hermitian or anti-Hermitian, this condition is generically satisfied since the spectrum is confined to lie entirely along the real or imaginary line (Fig.~\ref{fig1}e). However, when both $C_{ijmn}^e$ and $C^o_{ijmn}$ are nonzero, the spectrum can inhabit the full complex plane. 
Consequently, the spectrum, plotted as a function of $q_\perp$, need not retrace itself (Fig.~\ref{fig1}f). 
In this case, there exist segments in which no two extended Bloch modes have the same $\lambda$ and $q_\parallel$. Nonetheless, one can consider Bloch modes with complex wavenumbers $\tilde q_\perp = q_\perp +i \kappa (q_\perp)$. By analytically continuing $\hat D$ into the complex Brillouin zone, the eigenvalues can flow together to enable interference at the boundary.

Finally, we note that the use of the two specific moduli $G^e$ and $G^o$ in Eq.~(\ref{mat}) is purely illustrative. More generally, two necessary conditions must be met in order for the non-Hermitian skin effect to emerge within the continuum description of an elastic medium. First, there must be at least one non-vanishing modulus in both $C_{ijmn}^e$ and $C_{ijmn}^o$ in order for the spectrum to occupy the complex plane. Second, the system must be anisotropic in order for the spectrum not have a reflection symmetry over the system's boundary ($q_\perp \mapsto - q_\perp$).   
Finally,
we note that $\hat D$ has inversion symmetry, which implies that the skin modes occur in pairs localized to each boundary~\cite{Hofmann2020}. This is a contrast to systems in which an external medium enables an effective violation of Newton's third law~\cite{Ghatak2019Realization,Rosa2020,Zhou2020}. The continuum limit of these systems (whose interactions do not conserve linear momentum) yields $\hat D \propto q$, rather than the $\hat D \propto q^2$ dependence characteristic of elasticity (see S.I.).

\begin{figure}[t!]
    \includegraphics[width=0.5\textwidth]{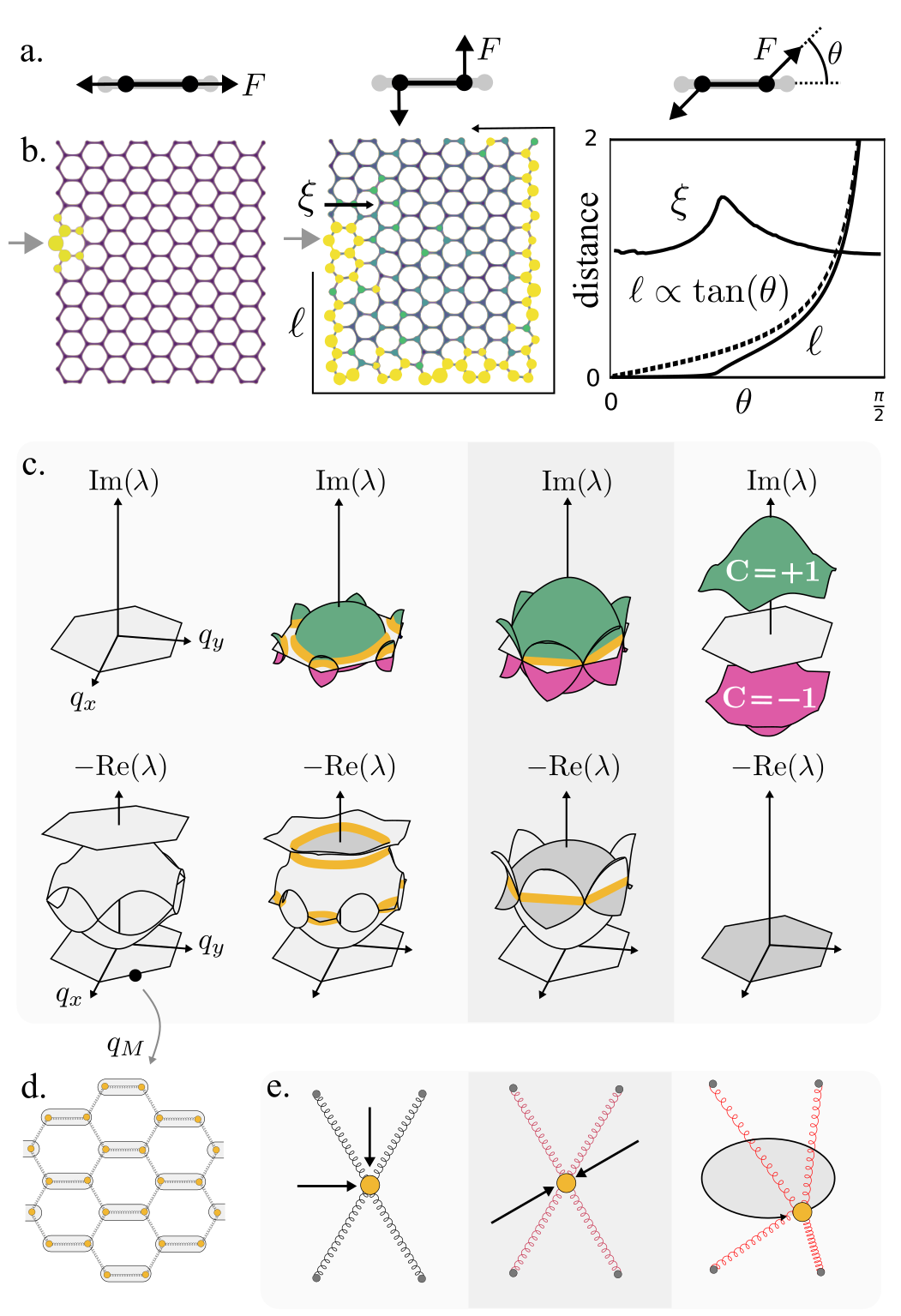}
    \caption{
    {\bf Non-Hermitian topological transition and exceptional rings.} {\bf a.}~The generalized Hookean spring with the force $\vb F$ oriented at an angle $\theta$ with respect to the bond vector. {\bf b.}~Simulations with $\theta=0$ (left) and $\theta=\pi/2$ (right) in which a particle at the edge is vibrated. Two lengths, the penetration depth $\xi$ and the propagation distance $\ell$, are plotted as a function of $\theta$. {\bf c.}~The spectrum plotted over the Brillouin zone for $\theta=0, \pi/12, \pi/6, \pi/2$. Regions of positive (green) and negative (purple) Berry curvature are highlighted. The orange lines denote exceptional rings. {\bf d.}~At point $\vb q_M$, pairs of masses move in tandem. {\bf e.}~The eigenmodes (arrows) of the effective single particle system at $ \theta=0,\pi/6, \pi/2$. 
    }
    \label{fig4}
\end{figure}

\emph{Microscopic model\textemdash} A ubiquitous minimal model for elastic solids is a collection of masses connected by Hookean springs~\cite{Paulose2015,Paulose2015a,Rocklin2017,Chen2014,Guest2003,Rocklin2017,Kane2014, Fruchart2020}. The Hookean spring captures two generic features of elasticity. First, the interaction conserves linear momentum, since the forces on the two participating particles are equal and opposite. Second, the force only depends on the change in bond length. Hence, the emerging mechanical response will be sensitive only to intrinsic changes in geometry. Yet, the Hookean spring has an additional feature built in: its force law follows from the gradient of a potential. Here, we retain the assumptions of length dependence and linear momentum conservation, and we study the most general 2D linear pairwise interaction when only energy conservation is lifted~\cite{Scheibner2020}:
\begin{align}
    \vb F(r) = - (k \vb{\hat r} + k^a \boldsymbol{\hat{\phi}}) \delta r  \label{force},
\end{align}
where $\vb{\hat r}$ ($\boldsymbol{\hat{\phi}}$) is a unit vector pointing along (transverse to) the bond vector, $\delta r$ is the change in length of the bond, and $k$ and $k^a$ are spring constants (see Fig.~\ref{fig4}a). 

When the bond is taken on a closed cycle, the work done $W = \oint \vb F \vdot \dd \vb r$ is equal to  $k^a$ times the area enclosed by the path. Hence, when $k^a \neq 0$,  Eq.~(\ref{force}) cannot be derived from a potential. 
In principle, Eq.~(\ref{force}) can be paired with any form of dynamics that governs the temporal evolution of the system. Here, for concreteness, we will interpret our results in the context of an overdamped equation of motion: $\Gamma \partial_t \vb u = \vb F$, where $\vb u$ is the displacement of the particle and $\Gamma$ is a drag coefficient. By adjusting the angle $\theta= \arctan(k^a/k)$ between $\hat {\vb r}$ and $\vb F$, we can interpolate between longitudinal and transverse interactions~\cite{Wiegmann2014,vanZuiden2016,Han2020,Banerjee2017,Galda2016}.
In the S.I., we show how the equations governing the \emph{overdamped} dynamics of an active solid with  \emph{non-conservative} bonds described by Eq.~(\ref{force}) are the same as those governing the {\it inertial} dynamics of a gyroscopic metamaterial~\cite{Nash2015,Wang2015} with \emph{conservative} spring-like interactions and weak dissipation. 
However, we note that deformation cycles performed with the gyroscopic media do not extract energy since the left hand side of Eq.~(\ref{force}) represents torques, not forces. Similarly, in the continuum treatment of gyroscopes, the left hand side of Eq.~(\ref{mat}) represents angular momentum currents, not stresses.

\emph{Generalized $\PT$ symmetry and energy cycles\textemdash}~For a generic network of masses connected by the bonds in Eq.~(\ref{force}), the linear relationship between forces $\vb F(\vb x)$ and displacements $\vb u (\vb x)$ can be captured by a dynamical matrix formalism $\vb F(\vb x) = \sum_{\vb x'} D(\vb x, \vb x') \vb u (\vb x')$, where $\vb x$ and $\vb x'$ are lattice sites and $D(\vb x, \vb x')$ is the dynamical matrix. The mere fact that the forces and displacements are real implies that $\comm{\K}{D(\vb x, \vb x')}=0$, where $\K$ is complex conjugation. Since $\K$ is an anti-unitary operator with $\K^2=1$, we say that the dynamical matrix has a generalized $\PT$ symmetry~\cite{Mostafazadeh2015,ashida2020,Bender1998,Fruchart2020phase}, see S.I. 

The $\PT$ symmetry has the following physical consequence: if a given eigenvalue $\lambda$ of $D$ is real, then the corresponding eigenvector $\vb u_\lambda (\vb x)$ may be chosen real. Since $ \vb u_\lambda(\vb x)$ is real, the corresponding trajectory of each particle traces out straight lines in time (Fig.~\ref{fig4}e left). Moreover, non-real eigenvalues come in complex conjugate pairs $\lambda_{\pm} = \lambda_R \pm i \lambda_I$ with eigenvectors of the form:
\begin{align}
    \vb u_{\lambda_{\pm}} (\vb x) = \vb v(\vb x) \pm i \vb w(\vb x),
\end{align}
where $\vb v(\vb x)$ and $\vb w (\vb x)$ are real vectors. Physically, a complex eigenvalue indicates energetic gain or loss. In this case, the eigenmode cycles between two states $\vb v(\vb x)$ and $\vb w(\vb x)$. Since the bonds are non-potential, the cycles result in the injection (or removal) of energy (Fig.~\ref{fig4}e right). If all the eigenvalues of $D$ are real, we say that $D$ is $\PT$-unbroken, and $\PT$-broken otherwise.

\emph{Non-Hermitian topological transition\textemdash}~Given a microscopic model, we can study not only the acoustic bands (accessible within the continuum theory) but also features of the optical bands. Figure~\ref{fig4}b shows the response of a honeycomb lattice to vibrations applied at the boundary in two extreme cases: the passive Hookean limit $\theta=0$, and the active transverse limit  $\theta=\pi/2$. In the transverse limit, we see the emergence of a sustained, unidirectional edge wave characteristic of a Chern insulator~\cite{Nash2015,Wang2015}. Due to the translation symmetry, we may express the dynamical matrix in terms of wave number $\vb q$ (see S.I.):
\begin{align}
    D_\theta (\vb q) = \cos (\theta) D_0 (\vb q) +\sin(\theta) D_{\pi/2} (\vb q), 
\end{align}
For Hermitian systems, nontrivial topology requires breaking time reversal symmetry (TRS): $D_\theta (- \vb q) = D^*_\theta(\vb q)$. However, here $D_\theta (\vb q)$ naively obeys TRS for all $\theta$ since the forces and displacements are real quantities. Nonetheless, band topology is still possible due to the violation of Hermiticity. 
At $\theta = \pi/2$, the dynamical matrix $D_{\pi/2} (\vb q)$ (restricted to its optical bands, see S.I.) is anti-Hermitian. Hence the relevant Hermitian Hamiltonian $H(\vb q)=i D_{\pi/2} (\vb q)$ violates TRS as required. When $\theta =\pi/2$, the lattice boundary hosts a chiral edge state due to the non-vanishing Chern numbers of the optical bands.

The transition between Hermitian and anti-Hermitian is accompanied by two length scales $\xi$ and $\ell$ (Fig.~\ref{fig4}b right). The first length scale $\xi$ is the penetration depth into the medium, which is set by the structure of the \emph{eigenvectors} of $D_\theta$. Like the Haldane model~\cite{Haldane1988}, $\xi$ is roughly constant as the gap opens and closes. Yet, the edge modes do not become visible until large values of $\theta$ are probed. This effect can be traced to the second length scale $\xi$, which is the distance the wave propagates around the edge. This length scale is set by \emph{eigenvalues} of the dynamical matrix. For a given mode,  $\ell \approx \tau \omega /q$, where $\tau=-1/\Re(\lambda)$ is the decay rate and $\omega=\Im(\lambda)$ is the oscillation frequency. Hence, the localized edge mode becomes apparent close to the anti-Hermitian limit, i.e. $\theta$ near $\pi/2$.  

\emph{Mechanical exceptional points\textemdash}~Insight into the transition is gained by examining the point $\vb q_M$ in the Brillouin zone. At this point, the $4 \times 4$ dynamical matrix $D(\vb q_M)$ can be reduced to an effective dynamical matrix that governs the motion of a single particle in a trap (Fig.~\ref{fig4}d-e):
\begin{align}
    D_{\text{eff}} =  -\mqty( \cos \theta & -3 \sin \theta \\ \sin \theta & 3 \cos \theta  ),
\end{align}
with eigenvalues $\lambda_\pm = -2 \cos \theta \pm \sqrt{2 \cos(2\theta)-1}$, see S.I. For $\theta < \pi/6$, two real modes exist which trace out straight lines (Fig.~\ref{fig4}e left); for $\theta> \pi/6$, the eigenmodes trace out cycles (right). At the transition $\theta=\pi/6$, $D_{\text{eff}}$ permits only a single eigenvalue $\lambda_- = \lambda_+$. However, the anisotropy of the trap and the chirality of the bonds imply that no two linear eigenmodes can have the same eigenvalue unless they are parallel. Hence, the two independent modes coalesce, indicating that the dynamical matrix is defective (i.e., non-diagonalizable). Such occurrences, known as \emph{exceptional points}, are generic features of transitions between $\PT$-broken and $\PT$-unbroken phases~\cite{Mostafazadeh2015,Kawabata2019PRL,Zhou2019,Xiong2018,Ghatak2019,Heiss2012,Lau2018}. Here, the exceptional points take on a clear physical meaning: they mark the crossovers between eigenmodes with linear motion, and eigenmodes with circular motion necessary to sustain active waves. 

For the honeycomb lattice, the exceptional points do not merely occur at point $\vb q_M$. Rather, they occur along 1D rings depicted in orange in Fig.~\ref{fig4}c.
In the S.I. we show that the inversion symmetry of the honeycomb lattice gives rise to a second manifestation of $\PT$ symmetry, given by $\PT = \K U$, where $U = 1 \otimes \sigma_x$ acts on $D_\theta(\vb q)$. This $\PT$-symmetry applies locally at each point in the Brillouin zone. Hence contiguous regions of the Brillouin zone form $\PT$-broken and $\PT$-unbroken phases bounded by rings of exceptional points~\cite{Yoshida2019qu,Yoshida2019cl,Okugawa2019,Budich2019}.

\begin{figure}[t!]
    \includegraphics[width=0.5\textwidth]{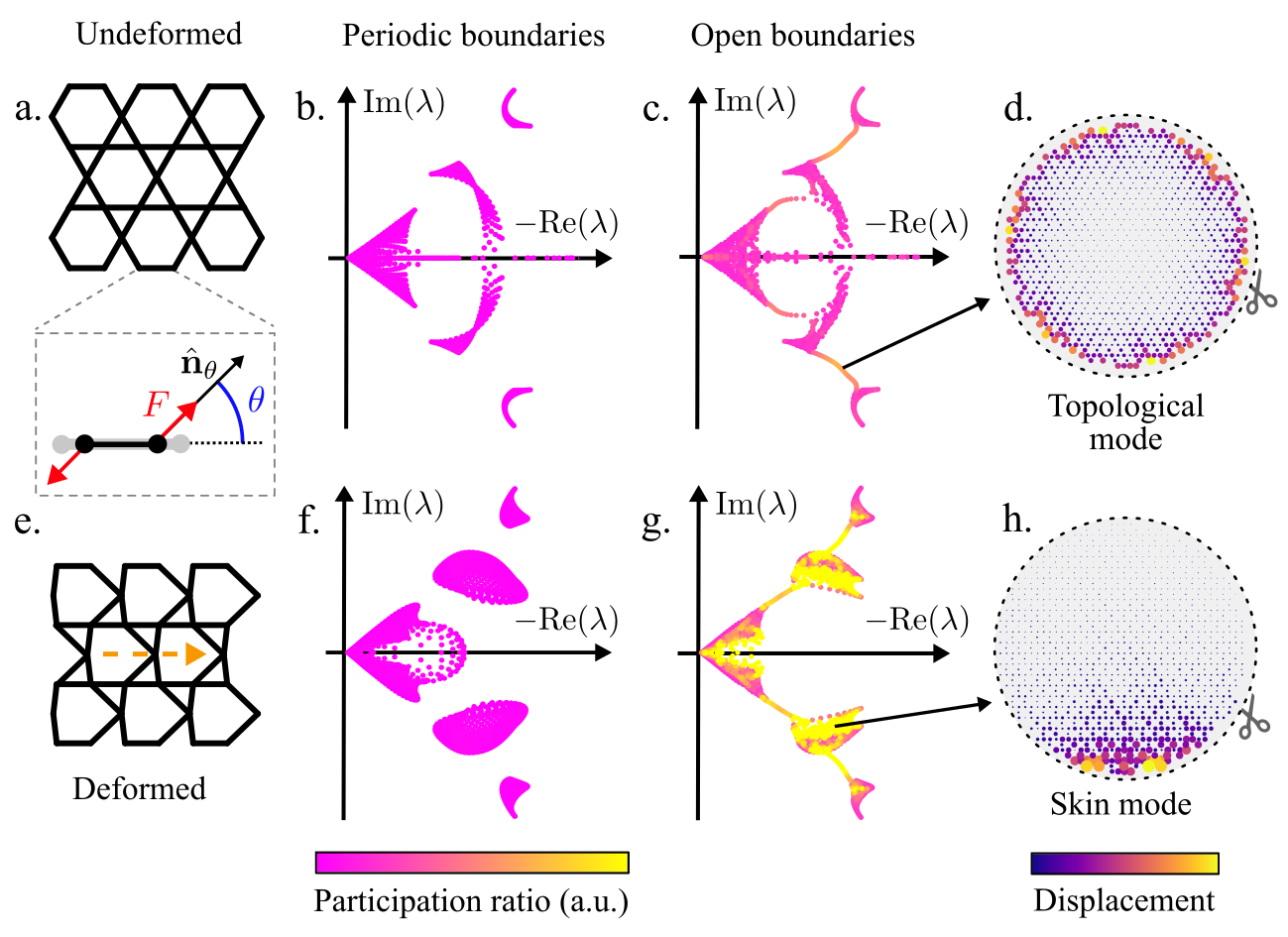}
    \caption{
    {\bf Microscopic model and skin effect.} 
    {\bf a.~} An undeformed kagome lattice with generalized Hookean springs. Inset: the force law displayed in Eq.~(\ref{force}). {\bf b.}~The spectrum when computed on a system with periodic boundaries. {\bf c.}~The spectrum when computed on a circle with pinned boundaries.  {\bf d.}~A set topological edge states connects distinct bands. The eigenmode of an edge state with displacements visualized by size and color. {\bf e.}~A deformed kagome lattice. Due to the chirality in the spring, deformation breaks reflection over horizontal axis. {\bf f.}~The periodic spectrum. {\bf g.}~The spectrum with open boundaries. {\bf h.}~Visualization of the non-Hermitian skin mode. 
    }
    \label{fig3}
\end{figure}

\emph{Non-Hermitian edge modes\textemdash}
Finally, we note that the skin effect, previously discussed in the long-wavelength limit within continuum theory, has a counter part in the optical bands of the lattice models. In Fig.~\ref{fig3}, we place the active bonds on an undeformed kagome lattice illustrated in panel (a) and compute the spectrum with periodic (b) and open (c) boundaries.
We color the modes by their participation ratio $\sum_{\vb x} \abs{\vb u (\vb x)}^4$, which serves as  a proxy for localization~\cite{Vitelli2010}.
For the undeformed kagome lattice, the sole difference between the periodic and open boundary spectra is the presence of a sub-extensive number of localized topological modes that span the band gaps. However, when we introduce a small deformation to the kagome lattice (Fig.~\ref{fig3}e-g), we observe a dramatic departure. 
We find that the open boundary system not only contains gap-spanning boundary modes, but 
the bulk bands become highly localized and their distribution in the complex plane changes dramatically. 
It is instructive to note the qualitative difference between the topological and skin modes (d and h). The localized bulk modes are confined to a single direction, whereas the topological modes are confined to all boundaries and decay into the bulk.

\emph{Conclusions\textemdash}~Our work brings to light the non-Hermitian phenomena that arise at the boundary of elastic media for which energetic sources (powered by internal activity or external fields) modify the relationship between static deformation and stress.

\begin{acknowledgments}
V.V.~was supported by the Complex Dynamics and
Systems Program of the Army Research Office under grant W911NF-19-1-0268. W.T.M.I. and V.V.~acknowledge support through the Chicago MRSEC, funded by the NSF through grant DMR-1420709. 
C.S.~was supported by the National Science Foundation Graduate Research Fellowship under Grant No. 1746045. W.T.M.I.~acknowledges support from NSF EFRI NewLAW grant 1741685 and NSF DMR 1905974. We would like to thank A. Souslov,  M. Fruchart, S. Huber, R. Thomale, and the anonymous referee for helpful discussions and useful suggestions. 
\end{acknowledgments}

\pagebreak

\section{Supplementary Information}

\setcounter{figure}{0}
\renewcommand{\thefigure}{{S\arabic{figure}}}

\setcounter{equation}{0}
\renewcommand{\theequation}{{S\arabic{equation}}}

\subsection{Elastic skin modes}

\subsubsection{Non-Hermitian continuum mechanics}

Here we place elasticity within the broader context of non-Hermitian continuum mechanics. 
We start from the following generic linear momentum conservation equation for a viscoelastic medium:
\begin{align}
    \rho \ddot u_i  =& \partial_j \sigma_{ij} + f^b_i, \label{eq:OV}
\end{align}
where $u_i (\vb x)$ is the displacement field and $\rho$ is the mass density. The right-hand side of Eq.~(\ref{eq:OV}) divides the total force density into two contributions: the divergence of the stress tensor $\partial_i \sigma_{ij}$ captures the internal forces that a solid exerts on itself without requiring contact with an external medium, while $f^b_i$ captures interactions that require contact with an external medium to supply linear momentum. 

Significant attention has been given to the role of $f^b_i$ in non-Hermitian solids. A few examples include terms of the following form:
\begin{align}
    f^b_i = \Gamma_{ij} \dot u_j + K_{ij} u_j+ K_{ijk} \partial_j u_k +\dots 
\end{align}
The first term $\Gamma_{ij} \dot u_j$, for example, has been linked to symmetry protected exceptional rings in the context of Coriolis forces,  substrate drag, and magnetic fields~\cite{benzoni2020,Yoshida2019cl}. The second term, $K_{ij}u_j$, generically arises in systems with pinning~\cite{Nash2015, Wang2015}. Finally, the third term, $K_{ijk} \partial_j u_k$ arises naturally in systems that contain effective pairwise interactions that violate Newton's third law~\cite{Beatus2006,Brandenbourger2019,Rosa2020,Zhou2020, Ghatak2019Realization}. 
It has been shown that this term can give rise to the non-Hermitian skin effect~\cite{Rosa2020, Zhou2020, Ghatak2019Realization}. In these cases, background mechanisms, such as robotic actuators~\cite{Ghatak2019Realization} or rotor systems~\cite{Zhou2020}, provide a linear momentum bias that intuitively localizes all the vibrational modes to one side of the system. 

It is intriguing to ask: what are the consequences of internal forces that are non-Hermitian yet respect linear momentum conservation? These contributions arise from $\sigma_{ij}$, and can be decomposed as
\begin{align}
    \sigma_{ij} = \sigma_{ij}^u + \sigma_{ij}^{\text{int}}.
\end{align}
Here, $\sigma_{ij}^{\text{int}}$ are the stresses that arise due to internal degrees of freedom, such as nematic~\cite{Prost2015, Souslov2017} or piezoelectric~\cite{White1962} stresses. By contrast, $\sigma_{ij}^u$ are stresses due to geometric deformations. In the limit of a low Deborah number~\cite{Lakes2009}, one can generically write these terms as  
\begin{align}
    \sigma_{ij}^u = (C_{ijmn} + \eta_{ijmn} \partial_t ) \partial_m u_n
\end{align}
The second term, $\eta_{ijmn}$, is the viscosity tensor, and it couples to the rate of deformation. 
Materials with broken time reversal symmetry can exhibit a dissipationless odd (or Hall) viscosity $\eta_{ijmn}^o = - \eta^o_{mnij}$, extensively explored in the context of fluids~\cite{Banerjee2017,Han2020, Souslov2020PRE,Wiegmann2014, Soni2018, Avron1998, Sone2019, Sone2019arxiv}. 
 
However, the first term $C_{ijmn}$ is the elastic modulus tensor and couples only to the static strain. 
In active systems, $C_{ijmn}$ can acquire an antisymmetric (or odd) part $C^o_{ijmn} =- C^o_{mnij}$, which introduces non-Hermiticity into the equations of motion~\cite{Scheibner2020}. The elasticity is the unique term that does not require external sources of linear momentum, rate-dependent forces, or auxiliary internal degrees of freedom.  
Hence, the conceptual limit of a purely elastic solid provides insight into the role played by non-Hermiticity in the relationship between stress and static deformation.

\subsubsection{Non-Hermitian skin effect}
Here, we show how the presence of $C^o_{ijmn}$ in conjunction with anisotropy can give rise to the non-Hermitian skin effect. Specifically, we consider the effect of $C^o_{ijmn}$ on the spectrum of the operator $D_{in} = C_{ijmn}\partial_j \partial_m$. We will work in two dimensions, for which it is useful to introduce the following basis of $2 \times 2$ matrices:
\begin{align}
    \tau^0 =& \mqty(1 & 0 \\ 0 & 1)  &&\text{Dilation/Pressure} \\ 
    \tau^1 =& \mqty(0 & -1 \\ 1 & 0 ) &&\text{Rotation/Torque}\\
    \tau^2 =& \mqty(1 & 0 \\ 0 & -1) &&\text{Shear Strain/Stress 1}\\ 
    \tau^3=&\mqty( 0 & 1 \\ 1 & 0) &&\text{Shear Strain/Stress 2} 
\end{align}
Using this basis, we can represent $C_{ijmn}$ as a $4 \times 4$ matrix:
\begin{align}
    C^{\alpha \beta} = \frac12 \tau^\alpha_{ij} \tau^\beta_{mn} C_{ijmn}.
\end{align}
Notice that the operation $C^{\alpha \beta} \mapsto C^{\beta \alpha}$ corresponds to $C_{ijmn} = C_{mnij}$. Hence, introducing antisymmetry into the matrix $C^{\alpha \beta}$ indicates that the solid's elastic response does not derive from a potential energy. Furthermore, one can impose the requirements $C^{1\alpha} =0$, which implies that the solid does not have an internal source of angular momentum, and $C^{\alpha 1}=0$, which implies that solid-body rotations do not induce stress~\cite{Scheibner2020}. While metamaterials can violate these properties~\cite{Lakes2001,Lakes2016,Banerjee2017}, they are not necessary ingredients for the non-Hermitian skin effect.  
In the main text, we consider a solid with the following moduli:
\begin{align}
    C^{\alpha \beta} =2 \mqty( 0 & 0 & 0 & G^e+G^o \\ 0 & 0 & 0 & 0 \\ 0 & 0 & \mu & 0 \\ G^e-G^o & 0 & 0 & \mu ) \label{mainC}
\end{align}

Notice that the modulus $G^o$ violates the symmetry of $C^{\alpha \beta}$, and therefore the major symmetry of the tensor $C_{ijmn}$. The modulus $\mu$ is the standard isotropic shear modulus. In the main text, we omit the second row and second column of Eq.~(\ref{mainC}) as they are entirely zero. In standard tensor notation, we have:
\begin{align}
    C_{ijmn} =& ( G^e+G^o ) \tau^0_{ij} \tau^3_{mn} + ( G^e-G^o ) \tau^3_{ij} \tau^0_{mn}+ \\
    & \mu (\tau^2_{ij}\tau^2_{mn}+\tau^3_{ij}\tau^3_{mn} ) \nonumber 
\end{align}
Working in reciprocal space, the operator $\hat D$ may be written as the following matrix:
\begin{align}
    \hat D = -\mqty(2G^e q_x q_y+ \mu q^2  & G^e q^2 + G^o (q_x^2 - q_y^2)  \\ G^e q^2 - G^o (q_x^2 - q_y^2) &  2G^e q_x q_y+ \mu q^2 ),
\end{align}
where $q^2 = q_x^2 + q_y^2$. We find that $\hat D$ has the following eigenvalues:
\begin{align}
    \lambda= -\mu q^2 - 2 G^e q_x q_y \pm \sqrt{ - [G^o (q_x^2-q_y^2) ]^2+[G^e q^2]^2 }
\end{align}
 Notice that the spectrum does not retrace itself in the complex plane as a function of $q_x$ at constant $q_y$ (cf. main-text Fig.~1).
 Hence, when a boundary is introduced parallel to the $y$-axis, no two Bloch modes are able to interfere to satisfy the given boundary conditions (such as displacement- or stress-free). Yet, even in the presence of a boundary, the system still has approximate translational symmetry, so modes with spatial variation of the form $e^{i x q_x} e^{- \kappa x}$ are candidates for approximate solutions. As illustrated by Fig.~1c, introducing nonzero $\kappa$ and analytically continuing $\hat D(\vb q)$ restores the spectral degeneracies necessary to satisfy the system's boundary conditions.

 Finally, we note that the skin effect is sensitive to the angle of the boundary. For example, if the boundary is rotated $45^\circ$, then the spectrum becomes:
\begin{align}
    \lambda_{\text{rot}}= -\mu q^2 - G^e ( q_\parallel^2 -q_\perp^2) \pm \sqrt{ - [2G^o q_\parallel q_\perp ]^2+[G^e q^2]^2 }, \label{rotSpec}
\end{align}
where $q_\parallel$ ($q_\perp$) are parallel (perpendicular) to the boundary. Notice that Eq.~(\ref{rotSpec}) is even
under $q_\perp \mapsto - q_\perp$, thereby suppressing the skin effect.

\subsection{Transverse interactions}
\subsubsection{Active bond}
In the main text, we present the interaction:
\begin{align}
    \vb F(r) = - (k \vb{\hat r} + k^a \boldsymbol{\hat{\phi}}) \delta r  \label{ForceLaw},
\end{align}
where $\vb{\hat r}$ ($\boldsymbol{\hat{\phi}}$) is a unit vector pointing along (transverse to) the bond vector, $\delta r$ is the change in length of the bond, and $k$ and $k^a$ are the passive and active spring constants, respectively. Equivalently, we may express this force law as:
\begin{align}
\vb F  = -k^t \hat {\vb n}_\theta \delta r, \label{form2}
\end{align}
where $k^t =\sqrt{k^2 + (k^a)^2 }$ is the stiffness, $\hat{\vb n}_\theta$ is the unit vector rotated by an angle $\theta = \arctan(k^a/k)$ with respect to the bond vector, and  $\delta r = r-r_0$ is the displacement from its equilibrium separation $r_0$, see Fig.~2a of the main text. As a matrix, we may write Eq.~(\ref{ForceLaw}) as:
\begin{align}
    \mqty(F_r \\ F_\phi ) = -\mqty( k  & 0 \\  k^a & 0 ) \mqty( u_r \\ u_\phi ), \label{mat}
\end{align}
where $\vb u = \vb r - \vb r_0$ is the displacement of the bond vector $\vb r$ from its rest configuration $\vb r_0$. We note that the $k^a$ violates the symmetry of the matrix in Eq.~(\ref{mat}), and hence violates Maxwell-Betti reciprocity at the level of a single bond. 
Here, we show that Eq.~(\ref{ForceLaw}) is the most general 2D pairwise interaction obeying the following conditions:
\begin{enumerate}[(i)]
    \item \label{dist} The interaction only depends on the distance between the particles.
    \item \label{momentum} The interaction conserves linear momentum.
    \item \label{linear} The interaction is linear in its displacements from an equilibrium length $r_0$. 
\end{enumerate}
For a 2D interaction obeying property (\ref{dist}), it is sufficient to specify the forces $\vb F_1(r)$ and $\vb F_2(r)$ on the two participating particles as a function of their separation $r$. Condition (\ref{momentum}) implies that $\vb F_1(r) = - \vb F_2(r)$, so we need only specify a single function $\vb F(r)$. By assumption (\ref{linear}), we may write $\vb F(r) = -\vb k \delta r$, where $\vb k$ is a 2D vector and $\delta r = r - r_0$ is the displacement from mechanical equilibrium. By parameterizing $\vb k$ as $\vb k= k^t \hat {\vb n}_\theta$, we obtain Eq.~(\ref{form2}). 

We will often interpret our results in the context of an overdamped equation of motion: $\Gamma \partial_t \vb u = \vb F$, which reads:
\begin{align}
    \Gamma \partial_t u_i = - k^t \hat n_{\theta i} \delta r.  \label{dynamic}
\end{align}
With these dynamics, the conditions of dynamic stability of a single spring is: 
\begin{align}
\Gamma k^t \cos(\theta) \ge 0.        
\end{align}
Assuming without loss of generality that $\Gamma, k^t >0$, the system is stable for $ \theta \in [-\pi/2,\pi/2]$.

\begin{figure}
    \centering
    \includegraphics[width=0.4\textwidth]{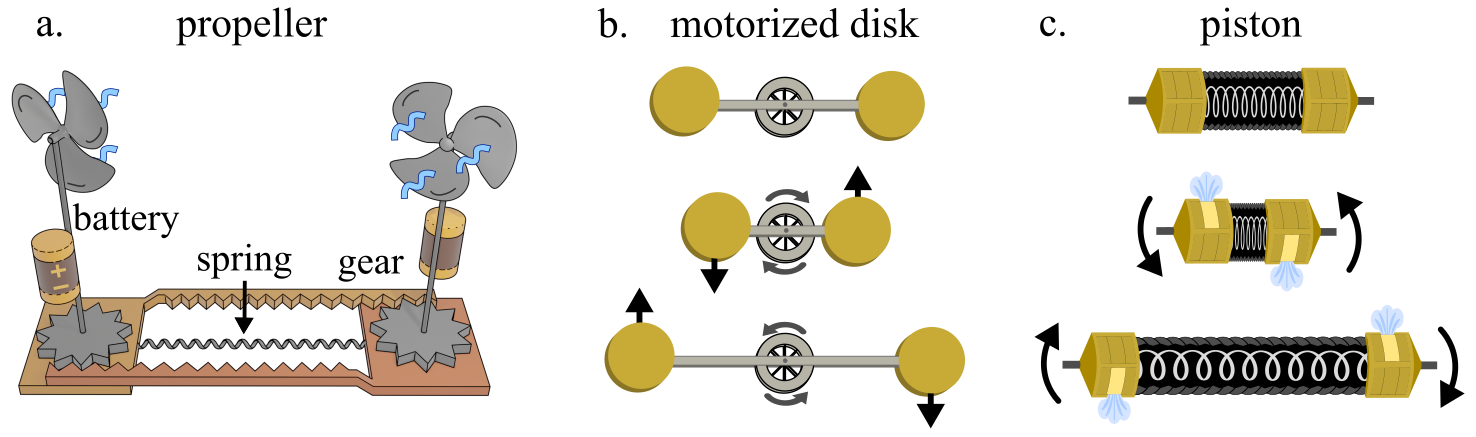}
    \caption{{\bf Mechanical realizations of active bonds.} {\bf a.}~Two battery-powered propellers rest on geared platforms. As the platforms slide together or apart, the propellers rotate to produce transverse forces. {\bf b.}~Two masses joined by an actuator with a motorized internal disk. The disk acts as a counterweight that converts internal angular momentum into orbital angular momentum of the masses. {\bf c.}~An internally pressurized piston. As the piston contracts or elongates, vents open to produce chiral forces.}
    \label{fig:realization}
\end{figure}

Figure~\ref{fig:realization} shows three simple mechanical realizations of the active bond adapted from Ref.~\cite{Scheibner2020}. The first example (a) consists of battery powered propellers that blow air at a constant rate. As the mechanism elongates and contracts, a gear system rotates the propellers to produce transverse forces which are proportional to the elongation or contraction for small strains. In example (b), two masses (yellow) are connected by a robotic bond housing an internal motor. As the bond elongates or contracts, the motor accelerates a disk (silver) whose reaction torques are felt by the masses. This example has the intriguing feature that the angular momentum is drawn from a source (i.e. the disk) housed inside the bond rather than a substrate or background fluid. The final example is a system of internally pressurized pistons. When the piston elongates or contracts, vents open to eject the internal fluid and thereby produce a transverse force.

\subsubsection{Mapping to dissipative gyroscopic metamaterials}
Now we show how Eq.~(\ref{dynamic}), which describes the \emph{overdamped} dynamics of a system with active, \emph{non-conservative} bonds, can be mapped onto the \emph{inertial} dynamics of a gyroscopic metamaterial with \emph{conservative} spring-like interactions and weak dissipation. Consider a system of gyroscopes suspended vertically with their bases fixed and free ends connected by Hookean springs, similar to those studied in Refs.~\cite{Mitchell2018,Mitchell2018Realization,Mitchell2018Nature,Nash2015, Wang2015}. 
In the fast-spinning limit, the angular momentum of the gyroscope is given by $\vb L =\Omega I \vecell$, where $\Omega$ and $I$ are, respectively, the angular velocity and moment of inertia about the axis of the gyroscope $\vecell$. When a force $\vb F$ is applied to the free tip, the gyroscope experiences a torque
\begin{align}
    \boldsymbol{\tau} =  \ell \vb F \cross \vecell
\end{align}
about its fixed base, where $\ell$ is the length of the gyroscope. We have assumed for simplicity that the center of mass of the gyroscope coincides with the point at which the force acts.  
Hence, we obtain the equation of motion:
\begin{align}
\Omega I \partial_t \vecell =& \partial_t \vb L \\
                            =& \boldsymbol{\tau} \\
                            =& \ell \vb F \cross \vecell. \label{Torque}
\end{align}
We will now use a Cartesian coordinate system in which $\hat {\bf z}$ is oriented vertically, parallel to the rest orientation of the gyroscopes. 
For small deflections from vertical, we may write the in-plane displacements of each gyroscope's tip as $u_x = \ell \hat \ell_x$ and $u_y = \ell \hat \ell_y$. 
When two gyroscopes are connected by Hookean springs, such that the spring is at its rest length $r_0$ when the gyroscopes are vertical, the in-plane spring forces are given by:
\begin{align}
F_i^{s} =  -k^t \hat n_i \delta r, \label{inplane}
\end{align}
to leading order in $u_i$. In Eq.~(\ref{inplane}), the index $i$ runs over $x$ and $y$, $\hat n_i$ is the in-plane bond vector, and $\delta r = r-r_0$ is the change in spring length. We note that geometric nonlinearities, such as out-of-plane motion of the spring, are subleading in $u_i$.

In addition to the spring force, we consider a dissipative drag term that takes the form:
\begin{align}
F_i^{d} = - \gamma  \partial_t  u_i. \label{drag}
\end{align}
Combining Eqs.~(\ref{Torque}-\ref{drag}), the in-plane equations of motion to leading order in $u_i$ become:
\begin{align}
    \frac{\Omega I}\ell \partial_t  u_i =- k^t \epsilon_{ij} \hat n_{j} \delta r - \gamma \epsilon_{ij} \partial_t u_j. \label{EOM1}
\end{align}
Now we define
\begin{align}
    \Gamma=&\sqrt{\qty(\frac{\Omega I}{\ell})^2 +\gamma^2} \\
    \hat n_{\theta i} =&  \qty( \frac{\Omega I}{\Gamma\ell} \epsilon_{ik} + \frac\gamma\Gamma \delta_{ik}  ) \hat n_k. \label{ndef}
\end{align}
We note that $\hat  n_{\theta i}$ in Eq.~(\ref{ndef}) is simply the vector $\hat  n_i$ rotated through the angle $\theta= \arctan( \frac{\Omega I}{\gamma \ell} )$. 
Using the definitions of $\Gamma$ and $\hat n_{\theta i}$, Eq.~(\ref{EOM1}) takes the form:
\begin{align}
\Gamma \partial_t u_i = -k^t \hat n_{\theta i} \delta r, \label{gresult}
\end{align}
which is mathematically identical in form to Eq.~(\ref{dynamic}). 
Notice, however, that Eq.~(\ref{gresult}) describes a
system of gyroscopes with \emph{conservative} interactions and weak dissipation. While Eq.~(\ref{gresult}) is mathematically equivalent to Eq.~(\ref{dynamic}), the righthand side of Eq.~(\ref{gresult}) should \emph{not} be interpreted as a physical force. Rather, the righthand side is a subset of the total torque that governs the evolution of the angular momentum parameterized by $u_i$. We note that when the dissipation vanishes, $\gamma = 0$,  equation Eq.~(\ref{gresult}) goes to the $\theta = \pi/2$ limit and describes a system for which energy is conserved.

\subsection{Exceptional points and rings}

\subsubsection{Generalized $\PT$ symmetry}

Here, we clarify our definitions and notation regarding $\PT$ symmetry~\cite{ashida2020, Mostafazadeh2015,Bernard2002, Bender1998}. Given a matrix $H$, we say that $H$ is $\PT$ symmetric if there exists an anti-unitary operator $\PT$  with $(\PT)^2=1$ such that $\comm{\PT}{H}=0$. We note that $\PT$ symmetry implies that all the eigenvalues of $H$ are either real or come in complex conjugate pairs. To see this, suppose
$\vb u$ is an eigenvector of $H$ with eigenvalue $\lambda$, and let $\vb v = \PT \vb u$. Then we have
\begin{align}
    H\vb v  =& H \PT \vb u \\
    =& \PT  H \vb u \\
    =& \PT \lambda \vb u \\
    =& \lambda^* \PT \vb u \\
    =& \lambda^* \vb v
\end{align}
Hence, $\vb v$ is an eigenvector of $H$ with eigenvalue $\lambda^*$. If all the eigenvalues of $H$ are real, then it is possible to simultaneously diagonalize $H$ and $\PT$. In this case, we say that $H$ is $\PT$-unbroken. If not all the eigenvalues of $H$ are real, we say that $H$ is $\PT$-broken. 

The transition between $\PT$-broken and $\PT$-unbroken phases are generically accompanied by exceptional points. To see why, first suppose that $H$ is a function of some parameter $\gamma$ such that an eigenvalue $\lambda$ is real for $\gamma \le 0$ and complex for $\gamma >0$. At $\gamma=0$, $H$ must experience a spectral degeneracy, since the complex conjugate pairs $\lambda$ and $\lambda^*$ in the broken phase must approach each other. For all $\gamma >0$, let $P$ be the projection operator onto the subspace of the eigenvectors $\vb u$ and $\vb v$ corresponding to eigenvalues $\lambda$ and $\lambda^*$ and related by $\vb v = \PT \vb u$. It is easily checked that $\PT P \PT = P$. Hence, $P \PT P$ is a $\PT$ symmetry of $PHP$. Thus, by choosing an appropriate basis on the subspace of $\vb u$ and $\vb v$, we may write $P \PT P = \K  $.  
Hence, in this basis, the most general form of $PHP$ as a $2\times2$ matrix is:
\begin{align}
    PHP = \mqty( a +b & c + d \\ c-d & a-b  )
\end{align}
whose eigenvalues are $\lambda_{\pm} = a \pm \sqrt{b^2+c^2 - d^2}$ and whose eigenvectors are:
\begin{align}
\vb w_{\pm} = \mqty( b \pm \sqrt{b^2 + c^2 - d^2 } \\ c-d )    
\end{align}
The degeneracy occurs at $d^2 = c^2 + b^2$, at which point the two eigenvectors $\vb w_+$ and $\vb w_-$ generically coalesce, indicating that $H$ exhibits and exceptional point. However, an exceptional point may be avoided if $H$ is fine tuned such that $b,c,d \to 0$ as $\gamma \to 0$.

Finally, we note that $H$ is $\PT$ symmetric if and only if $H$ is pseudo-Hermitian~\cite{ashida2020, Mostafazadeh2015}. 
For an invertible matrix $\eta$, we say $H$ is $\eta$-pseudo-Hermitian iff $\eta H \eta^{-1} = H^\dagger$. We say that $H$ is pseudo-Hermitian iff there exists an invertible matrix $\eta$ such that $H$ is $\eta$-pseudo-Hermitian. 
It can be shown that $H$ is $\PT$-unbroken if and only if it is $\eta$-pseudo-Hermitian for some a positive definite invertible matrix $\eta$. Finally, we often refer to instances of $\PT$ symmetry as \emph{generalized} $\PT$ symmetry to indicate that they do not represent physical notions of time reversal or parity.

\subsubsection{Energy cycles}

In our work, we highlight two primary manifestation of $\PT$ symmetry. The first arises from the fact that the dynamical matrix $D(\vb x, \vb x')$ for a system of $N$ particles is a real $2N \times 2N$ matrix. Hence, $D(\vb x, \vb x')$ commutes with the complex conjugation operator $\K$. Since $\K$ is an anti-unitary operator with $\K^2 =1$, it satisfies the definition of a (generalized) $\PT$ symmetry. In the language of Hermitian quantum mechanics, this symmetry would typically be referred to as time-reversal symmetry. This notion of $\PT$ symmetry does not rely on details of the network geometry, such as periodic order.

To understand physical significance of the $\PT$-broken and unbroken phases, consider an eigenvalue $\lambda$ with eigenvector $\vb u (\vb x)$. Let us impose a displacement controlled protocol in which a the eigenvector is actuated in a periodic manner, with the physical displacement given by $\vb d(\vb x , t) = \Re[ e^{i\omega_0 t} \vb u(\vb x)]$, where $\omega_0$ is an external parameter. If $\lambda$ is real, $\vb u(\vb x)$ may be chosen real. In this case, the trajectory of each particle traces out a straight line $\vb d (\vb x, t) = \vb  u(\vb x) \cos (w_0 t) $. The work done over the cycle is given by:
\begin{align}
    W =& \sum_{\vb x} \int_0^{2\pi/\omega_0}  \dot{\vb d} (\vb x, t) \cdot  \vb F(\vb x, t) \dd t \\
     =& - \lambda \omega_0 \sum_{\vb x}  \abs{ \vb u(\vb x)}^2  \int_0^{2\pi/\omega_0} \sin(\omega_0 t)\cos(\omega_0 t) \dd t \\
     =&0
\end{align}
Hence, the bonds do zero net work along the cycle. However, if $\lambda = \lambda_R + i \lambda_I$ is complex,  the eigenvector $\vb u(\vb x)$ must be of the form: $\vb u( \vb x) =  \vb v(\vb x) + i \vb w(\vb x) $, where $\vb w(\vb x)$ and $\vb v (\vb x)$ are real and may be chosen such that $\sum_{\vb x} \vb v(\vb x) \cdot \vb w(\vb x) = 0$. 
Then the displacement of a single particle traces out open arcs: $\vb d(x,t) = \vb v (\vb x) \cos(\omega_0 t) + \vb w (\vb x) \sin(\omega_0 t). $ Hence, the relative coordinates of the non-potential bonds trace out circles which correspond to energy injected or drawn from their microscopic sources. The total work is given by:
\begin{align}
    W =& \sum_{\vb x} \int_0^{2\pi/\omega_0}  \dot{\vb  d} (\vb x, t) \cdot  \vb F(\vb x, t) \dd t \\
    =& - \omega_0 \lambda_I \sum_{\vb x}  \int_0^{2 \pi/\omega_0} \bigg[ \abs{\vb w (\vb x)}^2 \cos[2](\omega_0 t) +\\  &\abs{\vb v(\vb x)}^2 \sin[2](\omega_0 t) \bigg] \dd t \\
    =& - \lambda_I \sum_{\vb x} \qty( \abs{\vb w (\vb x)}^2+ \abs{\vb v(\vb x)}^2), 
\end{align}
and is manifestly nonzero. Notice that the cycle may be performed quasistically and that reversing the cycle reverses the flow of energy. Operating the cycles at a fixed frequency $\omega_0$ merely illustrates the flow of energy. If the system is endowed with a given equation of motion in time, the frequency becomes a function of the eigenvalue $\lambda$. For example, for an overdamped system $\Gamma \partial_t \vb u (\vb x) = \vb F(\vb x)$, a real $\lambda$ indicates exponential relaxation (Fig.~2e left), while imaginary $\lambda$ indicates circulation (Fig.~2e right) since the energy injection balances the dissipative drag. The connection between $\PT$ symmetry and eigenmodes is also evident in the continuum: when $G^o=0$, the displacement field traces out straight lines (Fig.~1d left); and for sufficiently large $G^o$, the displacement field traces out ellipses (Fig.~1d center and right).

\begin{figure}
    \centering
    \includegraphics{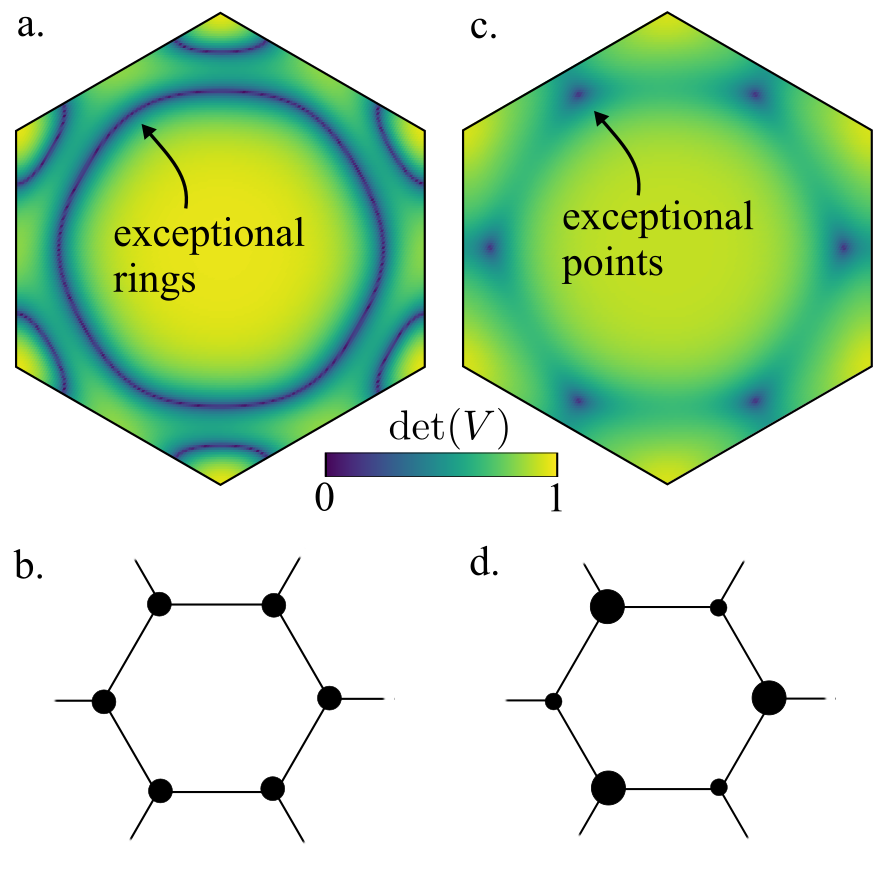}
    \caption{{\bf Exceptional rings and $\PT$ symmetry.}~{\bf a-b.}~The dynamical matrix for a honeycomb lattice with particle-hole symmetry has an emergent $\PT$ symmetry. We numerically diagonalize the dynamical matrix as $D (\vb q) = V(\vb q) E(\vb q) V^{-1}(\vb q) $, where $V$ is a matrix of normalized eigenvectors and $E$ is diagonal. Exceptional points correspond to $\det(V)=0$ and occur along exceptional rings. {\bf c-d.}~We break the particle-hole symmetry by introducing different drag coefficients (indicated by size) for the two particles in the unit cell. As a result, exceptional points emerge rather than exceptional rings. The examples shown correspond to active bonds of the form of Eq.~(\ref{ForceLaw}) with $\theta=\pi/12$. }
    \label{fig:ring}
\end{figure}

\subsubsection{Symmetry protected exceptional rings}
We now consider networks on periodic lattices, for which we may write the dynamical matrix as a $2n \times 2n$ matrix valued function of wavenumber $D(\vb q)$, where $n$ is the number of particles per unit cell. In this case, the $\PT = \K$ symmetry discussed in the preceding section can be expressed as $D(\vb q) = D^* (-\vb q)$. Hence, at a single point in $\vb q$-space, the $\PT=\K$ symmetry does not constrain the eigenvalues of the $2n \times 2n$ matrix $D(\vb q)$. However, as discussed below, the matrix $D(\vb q)$ itself can become $\PT$ symmetric when additional lattice symmetries are present.   

One of the consequences of the is second notion of $\PT$ symmetry is the emergence of symmetry protected exceptional rings (SPERs)~\cite{Yoshida2019cl,Yoshida2019qu,Okugawa2019,Budich2019}. 
Suppose that $D(\vb q)$ is $\PT$-broken at point $\vb q_0$ and $\PT$-unbroken at point $\vb q_1$. Let $\vb k_\alpha(\gamma)$ a family of homotopically equivalent curves in the Brillouin zone such that $ \vb k_\alpha(0) =\vb q_0$ and $\vb k_\alpha(1) =\vb q_1$.  For each $\alpha$, there must be at least one point $\gamma^*$ such that $D (\vb k_\alpha )$ transitions from being $\PT$-broken to $\PT$-unbroken, which generically corresponds to an exceptional point. By continuity, the the quantity $\vb k^*_\alpha$, treated as a function of $\alpha$, traces out a 1D curve in the Brillouin zone. Hence, the $\PT$ symmetry induces exceptional rings. If there are several values of $\gamma$ at which $D(\vb q)$ is defective, then there will be several exceptional rings traced out as a function of $\alpha$. 

In the main text, we present the example of the honeycomb lattice, whose dynamical matrix $D_\theta (\vb q)$ is given in Eq.~(\ref{dyn}). The honeycomb lattice has an inversion symmetry such that $U D_\theta (-\vb q) U^{-1} = D_\theta (\vb q)$, where 
\begin{align}
    U =  1 \otimes \sigma_x = \mqty( 0 & 0 & 1 & 0 \\ 0 & 0 & 0 & 1 \\ 1 & 0 & 0 & 0 \\ 0 & 1 & 0 & 0).
\end{align}
Combining the reality condition with the inversion symmetry, we can define the anti-unitary operator $\PT=U\K$. We note that $\comm{\PT}{D_\theta (\vb q)}=0$ for each point in $\vb q$ space. In Fig.~\ref{fig:ring}, we show the determinant of the matrix of right eigenvectors $V$ of $D_{\theta} (\vb q)$ over the Brillouin zone. The exceptional rings correspond to zeros of the determinant. When we remove the inversion symmetry, for instance by adding different masses or drag coefficients to the two particles, the exceptional vanish and only exceptional points remain.

In Ref.~\cite{Yoshida2019cl}, SPERSs are observed as a result of an emergent mechanical $\CP$ symmetry. We note that Ref.~\cite{Yoshida2019cl} utilizes an effective Hamiltonian formalism since the non-Hermiticty considered there enters through velocity dependent forces (e.g. Coriolis or drag forces) rather than the displacement-force relationship captured by the the dynamical matrix. While the physical origin of the non-Hermiticity and the mathematical object under consideration differ, our derivations follow closely those presented in Ref.~\cite{Yoshida2019cl}. This parallel arises since a matrix with $\PT$ symmetry can be transformed into one with $\CP$-symmetry through multiplication by $i$~\cite{Kawabata2019NatCom}.

\subsubsection{Consequences for Chern number}
We now show that violation of $\PT$ symmetry is necessary for a nonzero Chern number. 
Recall that the Berry curvature associated with the $n$th band of a Hermitian Hamiltonian of a 2D system is given by:
\begin{align}
    F (\vb q) = \epsilon_{ij} \pdv{A_j (\vb q)}{q_i}, \label{curve}
\end{align}
where $A_j(\vb q) $ is the Berry connection:
\begin{align}
    A_j (\vb q) = i\expval{\pdv{q_j}}{n(\vb q) },
\end{align}
and $\ket{n(\vb q)}$ is the eigenvector associated with the $n$th band~\cite{FRUCHART2013}. For a Hermitian operator $H$, the eigenvector $\ket{n(\vb q)}$ obeys the definition
\begin{align}
H \ket{n} = \lambda_n \ket{n}.
\end{align}
However, for a non-Hermitian operator, ambiguity arises in the definition of the eigenstate because the two vectors $\ket{n_R}$ and $\ket{n_L}$ defined by
\begin{align}
    H \ket{n^R} =& \lambda_n \ket{n^R} \\
    H^\dagger \ket{n^L} =& \lambda_n^* \ket{n^L}
\end{align}
need not be equal. Following the approach in, e.g.,~Ref.~\cite{Shen2018}, for a non-Hermitian system, one can define at least four different generalization of the Berry connection:
\begin{align}
    A_j^{RR} (\vb q) =& i\mel{n^R(\vb q)}{\pdv{q_j}}{n^R(\vb q) } \\
    A_j^{RL} (\vb q) =& i\mel{n^R(\vb q)}{\pdv{q_j}}{n^L(\vb q) } \\
    A_j^{LL} (\vb q) =& i\mel{n^L(\vb q)}{\pdv{q_j}}{n^L(\vb q) } \\
    A_j^{LR} (\vb q) =& i\mel{n^L(\vb q)}{\pdv{q_j}}{n^R(\vb q) }.
\end{align}
Generically, each of these definitions will give a different result for the local Berry curvature $F(\vb q)$. However, regardless of which expression for $A_i (\vb q) $ is used in Eq.~(\ref{curve}), the total Berry curvature of the band integrated over the full Brillouin zone (BZ) will be unchanged. Hence, the Chern number
\begin{align}
    C = \frac1{2\pi} \int_{BZ} \dd^2 q \, F(\vb q). 
\end{align} 
is insensitive to the choice of connection. In practice, we numerically compute the distribution of Berry curvature using the numerical approach described in Ref.~\cite{Fukui2013}.

First, let us suppose $H(\vb q) = H^*(- \vb q)$, which is the first notion of $\PT$ symmetry relevant for the dynamical matrices we study. If $H$ is $\PT$-unbroken, then the spectrum is entirely real and $\ket{n (-\vb q)} ^* =\ket{n (\vb q)}$. Therefore:
\begin{align}
    A_j^{RR} (\vb q) =& i\mel{n^R(\vb q)}{\pdv{q_j}}{n^R(\vb q) } \\
    =& i\mqty(\mel{n^R(-\vb q)}{\pdv{q_j}}{n^R(-\vb q) })^* \\
    =& - (A_j^{RR}(-\vb q))^*  \\
    =& - A_j^{RR} (-\vb q) , \label{Ccont}
\end{align}
where we have used the fact that the Berry connection is real. Notice that Eq.~(\ref{Ccont}) implies that the total Berry curvature vanishes over the full Brillouin zone, but may be locally nonzero. Hence, $H$ must have complex eigenvalues if it is to give rise to a nonzero Chern number. 

We can make even stronger statements given a $\PT$ symmetry such that $\comm{H(\vb q)}{\PT}=0$. If $H(\vb q)$ is $\PT$ unbroken, then for each eigenvalue $\lambda$ there exists an eigenvector $\ket{n^R (\vb q)}$ such that $\PT \ket{n^R (\vb q) } =\ket{n^R (\vb q) } $. Then we have:
\begin{align}
    A_j^{RR} (\vb q) =& i\mel{n^R(\vb q)}{\pdv{q_j}}{n^R(\vb q) } \\
    =& i\mel{\PT n^R(\vb q)}{\pdv{q_j}}{\PT n^R(\vb q) } \\
    =& i \mqty(\mel{n^R(\vb q)}{\pdv{q_j}}{n^R(\vb q) })^* \\
    =& - (A_j^{RR}(\vb q))^*  \\
    =& - A_j^{RR}(\vb q), \label{ACont}
\end{align}
where we have used the fact that the Berry connection is real. Equation~(\ref{ACont}) implies that the Berry curvature vanishes at all points such that $H(\vb q)$ is $\PT$ symmetric. This is illustrated in Fig.~2c in the main text, in which the Berry curvature (indicated by color) vanishes for all points for which the spectrum is real. Hence, $H$ must be $\PT$-broken in order to exhibit a nonzero Chern number.

\begin{figure}
    \centering
    \includegraphics[width=0.4\textwidth]{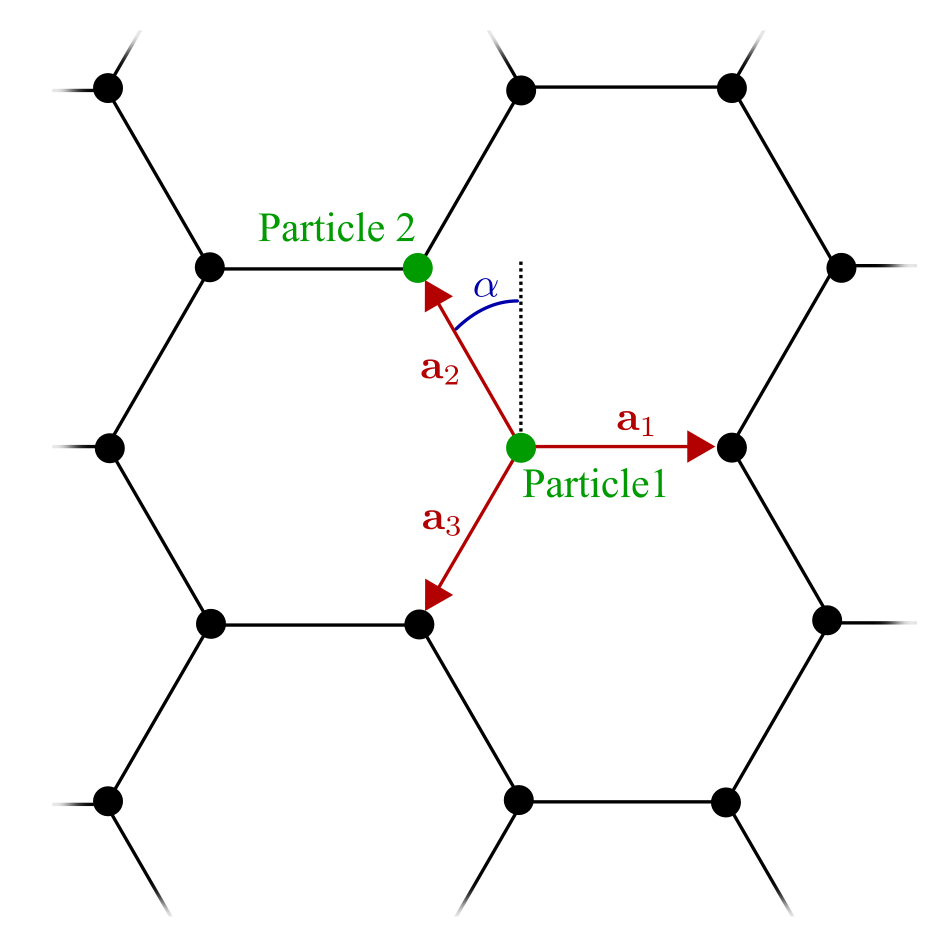}
    \caption{{\bf Notation for honeycomb lattice calculation.}~The unit cell consists of two particles, whose displacements are denoted by $\vb u_{1(2)}$ and $\vb F_{1(2)}$, respectively. }
    \label{fig:notation}
\end{figure}

\subsection{Honeycomb lattice} \label{spectrum}

In this section, we detail the calculations for the honeycomb lattice shown in Fig.~2. Throughout, we will rescale all lengths by the lattice spacing, and we will set $k^t=1$. Additionally, we will consider the family of deformed honeycomb lattices parameterized by the angle $\alpha$ shown in Fig.~\ref{fig:notation}.

\subsubsection{Expression for $D_\theta(\vb q)$}

As illustrated in Fig.~\ref{fig:notation}, the unit cell for the deformed honeycomb consists of two particles and three bond vectors given by:
\begin{align}
    \vb a_1=& a \mqty( 1 \\ 0 )\\
    \vb a_2=& a \mqty( -\sin \alpha  \\ \cos \alpha ) \\
    \vb a_3=&  a \mqty( -\sin \alpha \\ - \cos \alpha ),
\end{align}
where $a = \frac{1}{\sqrt3}$ is the bond length. 

Let $\vb u_1 (\vb x)$ and $\vb u_2 (\vb x)$ denote the displacements of particles 1 and 2, respectively, at position $\vb x$. The forces $\vb F_1 (\vb x)$ and $\vb F_2(\vb x)$ on these particles are given by  
\begin{align}
    \vb F_1 (\vb x) =& R(\theta) \sum_i A_i \qty[ \vb u_2 (\vb x + \vb a_i) - \vb u_1 (\vb x) ] \label{f1}\\ 
    \vb F_2 (\vb x) =& R(\theta) \sum_i A_i \qty[ \vb u_1 (\vb x - \vb a_i) - \vb u_2 (\vb x) ] \label{f2},
\end{align}
where
\begin{align}
    R(\theta) =& \mqty( \cos \theta & \sin \theta \\ -\sin \theta & \cos \theta ) \label{Rdef} \\
    A_1 =& \hat{\vb a}_1 \otimes \hat{\vb  a}_1 = \mqty(1 & 0 \\ 0 & 0 ) \\
    A_2 =& \hat{\vb a}_2 \otimes \hat{\vb  a}_2 = \mqty(\sin^2 \alpha & - \sin \alpha \cos \alpha \\  - \sin \alpha \cos \alpha & \cos^2 \alpha   )\\
    A_3 =& \hat{\vb a}_3 \otimes \hat{\vb  a}_3 =\mqty(\sin^2 \alpha &  \sin \alpha \cos \alpha \\   \sin \alpha \cos \alpha & \cos^2 \alpha   ) \label{Adef} 
\end{align}
The rotation matrix $R(\theta)$ encodes the transverse forces.   
Taking the Fourier transformation of Eqs.~(\ref{f1}-\ref{f2}) with the convention $f(\vb x) = \int \frac{\dd^2 q}{(2\pi)^2} f(q) e^{i \vb q \vdot \vb x}$ yields the following expression for $D_\theta(\vb q)$: 
\begin{align}
    D_\theta(\vb q) =  \mqty( -R(\theta) \sum_i A_i & R(\theta) \sum_i e^{ i \vb q \vdot \vb a_i} A_i \\
                        R(\theta) \sum_i e^{- i \vb q \vdot \vb a_i}A_i & -R(\theta) \sum_i A_i ). \label{dyn}
\end{align}
To obtain the eigenvalues of $D(\vb q)$, we solve the characteristic polynomial 
$P(\lambda) = \det[ D(\vb q) - \lambda I]$, 
where $I$ is the $4\times4$ identity matrix. If we assume an equation of motion of the form $\partial_t \vb u = \vb F$, and use the convention $f(t) = \int \frac{\dd \omega}{2\pi} f(\omega) e^{-i\omega t}$, the frequency of propagation associated with the eigenvalue $\lambda$ is given by $\omega = -\Im(\lambda)$, and the decay time $\tau$ is given by $1/\tau = -\Re(\lambda)$.

\subsubsection{Spectrum}

For the honeycomb lattice, we evaluate the characteristic polynomial at $\alpha = \pi/6$ and obtain:
\begin{align}
    p(\lambda)= \lambda\bigg\{& \lambda^3 + 6  \lambda^2 \cos \theta + \frac{\lambda}2\qty[18+ G(2 \cos 2\theta -1)] \nonumber \\ 
    &+ \frac32 G \cos \theta \bigg\}, \label{charpoly}
\end{align}
where all the $\vb q$-dependence is captured in the $\theta$-independent quantity:
\begin{align}
    G = 3-\cos q_y - 2 \cos \frac{q_y}2 \cos \frac{\sqrt3}{2}q_x.
\end{align}
Notice that the polynomial in Eq.~(\ref{charpoly}) has a trivial root $\lambda =0$ that corresponds to a zero mode. Furthermore, notice that the spectrum is symmetric under $q_y \mapsto -q_y$ and $q_x \mapsto -q_x$. Hence, when horizontal or vertical boundaries are introduced into the honeycomb lattice, the skin effect is suppressed due to this reflection symmetry.

\begin{figure}
    \centering
    \includegraphics[width=0.45 \textwidth]{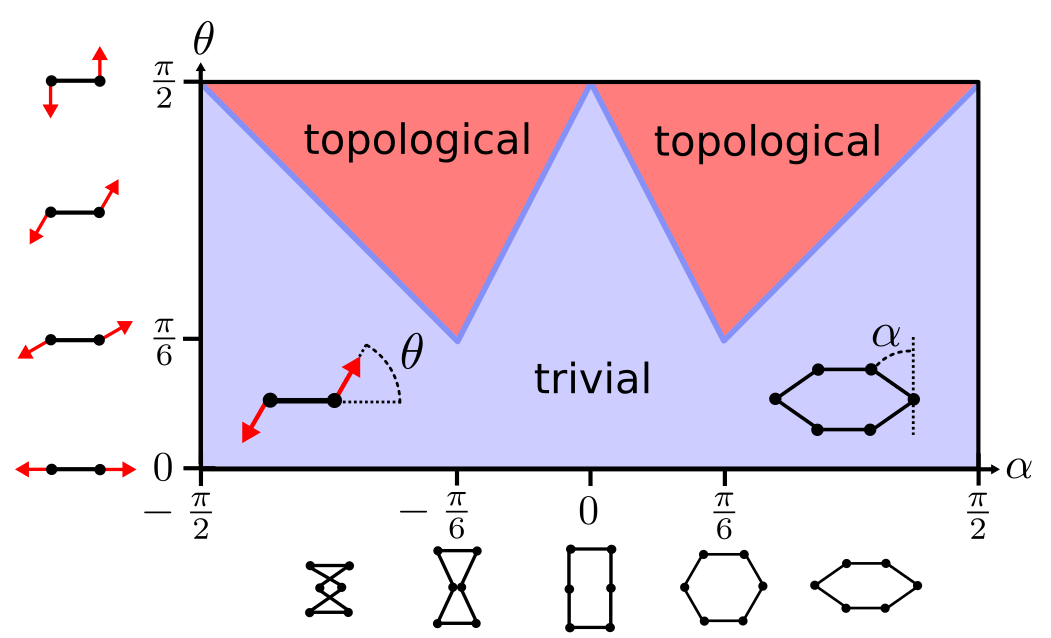}
    \caption{{\bf Phase diagram for deformed honeycomb lattice.} The angle $\alpha$ parameterizes the family of deformed honeycomb lattices. The angle $\theta$ characterizes the restoring force of the active bond. Red regions contain a topological gap while the blue regions are ungapped. }
    \label{fig:PhaseDiagram}
\end{figure}

\subsubsection{Gap opening}
Now we now compute the critical value of $\theta_T$ at which the gap opens for the family of deformed honeycomb lattices. The phase diagram is shown in Fig.~\ref{fig:PhaseDiagram}. For $\abs{\alpha} \le \pi/6$ we numerically observe that the band gap opens at the point:
\begin{align}
    \mqty(q_x \\ q_y )=\mqty( \frac{\sqrt 3 \pi}{1+\sin \alpha} \\ 0)  
\end{align}
and the points equivalent by symmetry. The roots of the characteristic polynomial at these points are given by:
\begin{align}
    \lambda =& 0 \\ 
    \lambda =& -i 2 \cos \theta \\ 
    \lambda =& -i 2 \cos \theta \pm i \sqrt{2 ( \cos 2\theta + \cos 4 \alpha) }.
\end{align}
Hence, we find that the topological transition occurs at
\begin{equation}
    \theta_T = 
    \begin{cases}
    \pi/2-2\alpha & 0 < \alpha \le  \pi/6 \\ 
    \pi/2+2 \alpha & - \pi/6 \le \alpha <0.
    \end{cases}
\end{equation}
Similarly, for $\pi/6 \le \abs{\alpha} \le \pi/2 $, we observe that the gap-opening occurs at 
\begin{align}
    \mqty(q_x \\ q_y )=&
    \mqty( \frac{\sqrt 3 \pi}{2(1+\sin \alpha)} \\  \frac{\sqrt 3 \pi }{ 2 \cos \alpha}),
\end{align}
and the points equivalent by symmetry. The roots at these points are given by:
\begin{align}
    \lambda =& 0 \\ 
    \lambda =& -i 2 \cos \theta \\ 
    \lambda =& -i 2 \cos \theta \pm i \sqrt{2 ( \cos 2 \theta - \cos 2 \alpha) },
\end{align}
which implies that the transition occurs at:
\begin{equation}
    \theta_T=
    \begin{cases}
    \alpha  & \pi/6 \le \alpha \le \pi/2 \\ 
    -\alpha & - \pi/2 \le \alpha \le -\pi/6.
    \end{cases}
\end{equation}

\subsubsection{Effective particle}
We now explain how the gap opening for the family of deformed honeycomb lattices corresponds to the effective particle shown in Fig.~2d-e. Since point $\vb q_M$ is equivalent to point $-\vb q_M$,  the matrix $D_\theta(\vb q_M)$ must be real. Using Eq.~(\ref{dyn}) for a honeycomb lattice ($\alpha =\pi/6$), we find the matrix to be of the form:
\begin{align}
    D_\theta(\vb q_M) = -\frac12 \mqty( 3R  & R [1- 2  \sigma_z] \\ R[1 - 2 \sigma_z] & 3R).
\end{align}
Notice that $D_\theta(\vb q_M)$ permits eigenvectors $(\vb u_1, \vb u_2)$ of the form $\vb u_1 = \vb u_2$, which correspond to the particles in a unit cell moving as a single mass. The effective dynamical matrix governing this equation is given by:
\begin{align}
    D_{\text{eff}} = -R[2-\sigma_z] = -\mqty( \cos \theta & -3 \sin \theta \\ \sin \theta & 3 \cos \theta  ), 
\end{align}
as referenced in the main text. The eigenvectors of $D_{\text{eff}}$ are given by:
\begin{align}
    \vb u_{\pm} = \mqty( -\cos \theta \pm \sqrt{2 \cos (2\theta) -1} \\ \sin \theta )
\end{align}
with eigenvalues $\lambda_{\pm} = -2\cos \theta \mp \sqrt{2 \cos(2\theta) -1}$. When $\theta=\pi/6$, the eigenvectors coalesce, giving rise to an exceptional point. 

A similar approach applies for the deformed honeycomb. For $\abs{\alpha} \le \pi/6$, the effective matrix is given by:
\begin{align}
D_{\text{eff}} = 4R\mqty( \sin[2](\alpha ) & 0 \\ 0 &  \cos[2](\alpha) ). 
\end{align}
For $\abs{\alpha}> \pi/6$, the effective dynamical matrix is given by:
\begin{align}
    D_{\text{eff}} = -2R\mqty(   [1+ \sin[2](\alpha)] & -  \cos \alpha \sin \alpha \\ -  \cos \alpha \sin \alpha &   \cos[2](\alpha) ).
\end{align}

\subsubsection{Comparison to the Haldane model}

\begin{figure}
    \centering
    \includegraphics[width=0.45 \textwidth]{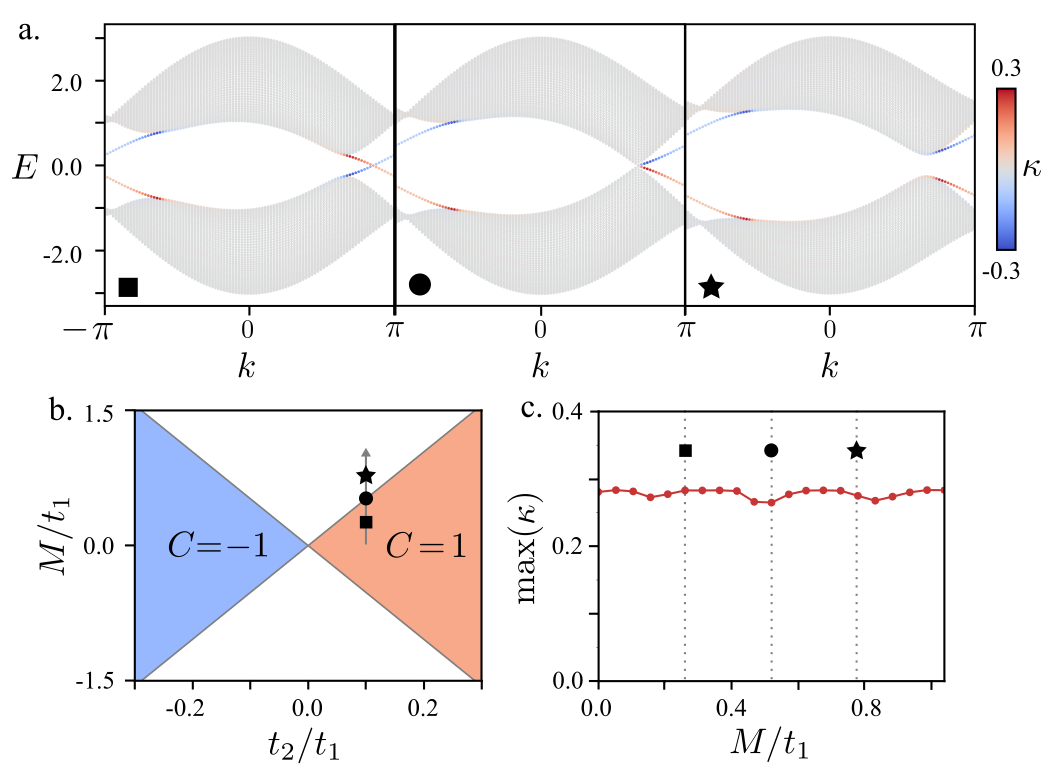}
    \caption{ {\bf Penetration depth in the Haldane model.} {\bf a.}~The spectrum for the Haldane model with ``zig-zag" termination is shown for three different values of parameters: $\{ M/t_1, t_2/t_1 \} = \{ 0.05 \times 3\sqrt{3},   0.1  \}$ (square),$\{ 0.1 \times 3\sqrt{3},   0.1  \}$ (circle), $\{ 0.15 \times 3\sqrt{3},   0.1  \}$ (star). The color indicates the  inverse penetration depth $\kappa$ (in units of the lattice spacing). {\bf b.}~The points in parameter space cross the boundary between a topological (red) and trivial (white) state. Here, $C$ denotes the Chern number of the top band. {\bf c.} The maximum localization for the spectrum is shown as a function of $M/t_1$ along a cut at constant $t_2/t_1=0.1$. We note that the penetration depth $\xi = 1/\kappa$ does not diverge as $M/t_1$ is tuned through the topological transition.  }
    \label{fig:Haldane}
\end{figure}

In this section, we compare and contrast the honeycomb lattice of active bonds to the Haldane model~\cite{Haldane1988}. 
The Hamiltonian for the Haldane model is given by:
\begin{align} \label{Haldane}
    H_{H} (\vb q) =& t_1 \mqty( 0 & \Delta (\vb q)  \\ \Delta^\dagger (\vb q) & 0 ) + M  \mqty( 1 & 0 \\ 0 &- 1) \\
    & + 2 t_2 \mqty(   \sum_i \sin( \vb q \vdot \vb b_i )& 0 \\ 0 & - \sum_i \sin ( \vb q \vdot \vb b_i )  ), \nonumber 
\end{align}
where $\vb b_i = \vb a_i - \vb a_{i+1}$, $t_1$ sets the nearest-neighbor hoppings, $M$ is the mass, and $t_2$ sets the next-nearest-neighbor hoppings, and $\Delta (\vb q) = \sum_i e^{i \vb q \vdot \vb a_i}$. We call the first term in Eq.~(\ref{Haldane}) $H_0$. We note that $H_0$ contains a chiral symmetry in the form of a sublattice symmetry $\sigma_z H_0 (\vb q) \sigma_z= - H_0 (\vb q) $. Furthermore $H_0$ contains a time-reversal symmetry $H_0^*(-\vb k) = H_0 (\vb k) $, which arises since the matrix $H_0$ is real in physical space. Finally, the combination of the chiral and time reveal symmetry imply a particle-hole symmetry $\sigma_z H^* (- \vb k) \sigma_z = - H(\vb k)$. Both the mass $M$ and the next-nearest-neighbor hopping $t_2$ violate the chiral (and hence the particle-hole symmetry) and ultimately give rise to band gaps. However, only the term $t_2$ violates time reversal symmetry, and hence makes the model compatible with nonzero Chern number. 

For the model in Eq.~(\ref{dyn}) with $\theta=0$ and $\alpha=\pi/6$ (corresponding to a honeycomb lattice with passive springs), we have \begin{align}
    D_0 (\vb q) = \mqty( -\frac32 & B(\vb q) \\ B^\dagger (\vb q) & - \frac32), 
\end{align}
where $B(\vb q) = \sum_i A_i e^{i \vb q \vdot \vb a_i }$. We note that the eigenvalues of  $B^\dagger B$ are $\frac94$ and $\abs{\Delta}^2$. Hence, there exist unitary matrices $U(\vb q)$ and $V(\vb q)$ such that 
\begin{align}
    B(\vb q) = U(\vb q) \mqty( \Delta(\vb q) & 0 \\ 0 & \frac32 ) V^\dagger (\vb q) . 
\end{align}
Consequently, we may write the dynamical matrix as:
\begin{align}
    D_0 = W  \mqty( 0 & 0 & 0 & \frac32 \\ 0 & 0 & \Delta  & 0 \\ 0 & \Delta^\dagger & 0 & 0 \\ \frac32 & 0 & 0 & 0 ) W^\dagger -\frac32 1_{4 \times 4}. \label{decomposed}
\end{align}
Where $W=\mqty( U(\vb q) \sigma_x & 0 \\ 0 & V (\vb q) )$ is a unitary matrix. Examining Eq.~(\ref{decomposed}), we find that the spectrum of $D_0$ is simply $\{ 0, \pm \abs{\Delta (\vb q)}  - \frac32, -3 \} $, which is equivalent to that of $H_0$ with the addition of two flat bands and an overall energy shift. 

Like $H_0$, the matrix $D_0$ contains a time reversal symmetry $D_0 (\vb q) = D_0^*(- \vb q)$ and a chiral symmetry arising from a sublattice symmetry (once the trace is removed). A nonzero $\theta$ introduces the perturbation 
\begin{align}
    D_\theta (\vb q) = \cos(\theta) D_0 ( \vb q ) +  \sin (\theta) D_{\pi/2} (\vb q),
\end{align}
where 
\begin{align}
    D_{\pi/2} (\vb q) = i\mqty( - \frac32 \sigma_y & \sigma_y B(\vb q) \\ \sigma_y B^\dagger (\vb q) &  -\frac32 \sigma_y ). 
\end{align}
Like $t_2$ and $M$ in the Haldane model, $D_{\pi/2} (\vb q)$ violates the sublattice symmetry. However, $D_{\pi/2}(\vb q)$ unlike $t_2$, respects the time-reversal symmetry. Nonetheless, $D_{\pi/2} (\vb q)$ is topologically nontrivial because it introduces non-Hermiticity. Indeed, when restricted to its nonzero modes, $D_{\pi/2}(\vb q)$ is anti-Hermitian, and on can define an effective Hamiltonian $H_{\text{eff}} (\vb q) = i D_{\pi/2} (\vb q)$ which is Hermitian and violates time reversal, sublattice, and chiral symmetry, thus enabling nontrivial topology.  Finally, we note that $D_\theta (\vb q)$ has well defined edge modes even when $\theta=0$. Like the Haldane model, these edge modes remain localized even as the system passes through a topological transition, as illustrated in Fig.~\ref{fig:Haldane}. While general divergent length scales, such as the Wannier correlation length~\cite{Chen2019}, have been identified with topological transitions, for the Haldane model and the transitions studied here, the penetration depth is on the order of the lattice spacing throughout the transition and is not set by the diverging length scale.

\subsection{Numerics}

\subsubsection{Elastic equations}
The normal modes in main-text Fig.~1 are generated by numerically solving the eigenvalue equation of the operator $D_{in} = C_{ijmn} \partial_j \partial_m$ on a domain with periodic boundaries in the $y$ direction and fixed boundaries in the $x$ direction. To do so, we take the Fourier transform of $\hat D$ in the $y$ direction, thereby introducing $q_y$ as an external parameter in the resulting 1D differential equation. We then solve the 1D equation by introducing a finite difference method. We represent:
\begin{align}
    \partial_x =& \frac12 \mqty(  0 & 1 & 0 & \cdots  & & -P \\ 
                                -1 & 0 & 1 &  & &  \\ 
                                0 & -1 & 0 &  & & \vdots  \\
                                \vdots  &    &  & \ddots & &  \\
                                  & & &   & 0& 1 \\
                                 P &  & \cdots  &  & -1 &0 )
\end{align}
and
\begin{align}
    \partial_x^2 =& \frac12 \mqty(  -2 & 1 & 0 & \cdots & & P \\ 
                                1 & -2 & 1 &  & &  \\ 
                                0 & 1 & -2 & & & \vdots  \\
                                \vdots & & & \ddots & &  \\
                                  & & & & -2 & 1 \\
                                 P & & \cdots & & 1 & -2 ),
\end{align} 
where $P=0$ for zero displacement boundaries and $P=1$ for periodic boundaries in the $y$ direction. Using these expressions, we can represent $D_{in}$ as a matrix and compute its eigenvalues and eigenvectors. For the data shown in Fig.~1, we descretize the $x$ axis into 1000 steps and take $\{ q_y, G^e, G^o, \mu \} = \{ 0.1, 1.0,0.0,2.0\}$ (left),  $ \{ 0.1, 0.0,1.7,0.0\}$ (center),  $ \{ 0.1, 1.0,1.7,2.0\}$ (right).

\subsubsection{Simulations}

The simulations shown in Fig.~2 are performed on a lattice of 15 by 15 unit cells with free boundaries on all sides. The position of each particle is integrated according the following second-order Runge-Kutta scheme:
\begin{align}
\vb x_i \qty(t + \frac12 \Delta t) =& \vb x_i (t) + \frac12 \Delta t \vb F_i \qty[ \underline{\vb x}(t)] \\
\vb x_i (t + \Delta t) =& \vb x_i (t) +  \Delta t \vb F_i \qty[ \underline{\vb x}\qty (t+\frac12 \Delta t)],
\end{align}
where $\Delta t$ is the time step, $\vb x_i(t)$ is the position of the $i$th particle, $\vb F_i$ is the force on the $i$th particle, and $\underline{\vb x }(t)$ is the list of all particle positions. In our simulations, time is measured in units of $\Gamma/k$ and distances are measured in units of the lattice spacing. 

The particles are initialized at their rest positions, and a single particle on the boundary of the lattice is subjected to controlled displacements given by $\vb u (t) = (A \cos(\omega t), 0)$. For the simulations shown, $A=0.1$, $\omega =0.5$, $\Delta t=0.1$, and $\theta$ takes the values $\{0, \pi/2\}$. The final time shown in the figure is $t_f = 100$ for $\theta=0$ and $t_f=50$ for $\theta= \pi/2$. 

\subsubsection{Characterization of length scales}
We now describe the numerical procedure for determining the length scales shown in Fig.~3. Using the notation of Fig.~\ref{fig:notation}, we consider a system with $N_x=30$ unit cells in the horizontal ($x$) direction and unbounded in the vertical ($y$) direction. Performing a Fourier transform with respect to the vertical coordinate gives rise to a $2N_x$ by $2N_x$ matrix of the form:
\begin{widetext}
\begin{align}
    D(q_y) = \begin{bmatrix}- \frac32 R  & R B  & & & & & \\
                        R B^\dagger  & - \frac32 R & R C & & & \mbox{\Huge0} & \\    
                            & R C^\dagger  & - \frac32 R & R B & & & \\
                            & & R B^\dagger & -\frac32 R & R C & &  \\
                            & & & R C^\dagger & \ddots & & \\
                            & \mbox{\Huge0} & & & &   \ddots & R B   \\
                            & & & & & R B^\dagger & - \frac32 R   \\
    \end{bmatrix},
\end{align}
\end{widetext}
where 
\begin{align}
   B=& A_2 + e^{i q_y} A_3 \\
   C=& e^{i \frac{q_y}2 } A_1,
\end{align}
with $A_1$, $A_2$, $A_3$ and $R$ defined in Eqs.~(\ref{Rdef}-\ref{Adef}). 

We numerically find the eigenvalues and eigenvectors of $D(q_y)$. We sort the eigenvectors by their localization, and we find that the edge modes are well characterized by the form $\vb u(\theta, q_y, x) \propto e^{- x / \xi(\theta, q_y)}$. In Fig.~3b, we plot $\xi$ and $\ell$ as a function of $\theta$ at $q_y = \pi$.


\begin{thebibliography}{10}
\expandafter\ifx\csname url\endcsname\relax
  \def\url#1{\texttt{#1}}\fi
\expandafter\ifx\csname urlprefix\endcsname\relax\def\urlprefix{URL }\fi
\providecommand{\bibinfo}[2]{#2}
\providecommand{\eprint}[2][]{\url{#2}}

\bibitem{Shmuel2020}
\bibinfo{author}{Shmuel, G.} \& \bibinfo{author}{Moiseyev, N.}
\newblock \bibinfo{title}{Linking scalar elastodynamics and non-hermitian
  quantum mechanics}.
\newblock \emph{\bibinfo{journal}{Phys. Rev. Applied}}
  \textbf{\bibinfo{volume}{13}}, \bibinfo{pages}{024074}
  (\bibinfo{year}{2020}).

\bibitem{Brandenbourger2019}
\bibinfo{author}{Brandenbourger, M.}, \bibinfo{author}{Locsin, X.},
  \bibinfo{author}{Lerner, E.} \& \bibinfo{author}{Coulais, C.}
\newblock \bibinfo{title}{Non-reciprocal robotic metamaterials}.
\newblock \emph{\bibinfo{journal}{Nature Communications}}
  \textbf{\bibinfo{volume}{10}}, \bibinfo{pages}{4608} (\bibinfo{year}{2019}).

\bibitem{Yoshida2019cl}
\bibinfo{author}{Yoshida, T.} \& \bibinfo{author}{Hatsugai, Y.}
\newblock \bibinfo{title}{Exceptional rings protected by emergent symmetry for
  mechanical systems}.
\newblock \emph{\bibinfo{journal}{Phys. Rev. B}}
  \textbf{\bibinfo{volume}{100}}, \bibinfo{pages}{054109}
  (\bibinfo{year}{2019}).

\bibitem{Rosa2020}
\bibinfo{author}{Nora~Rosa, M.~I.} \& \bibinfo{author}{Ruzzene, M.}
\newblock \bibinfo{title}{Dynamics and topology of non-hermitian elastic
  lattices with non-local feedback control interactions}.
\newblock \emph{\bibinfo{journal}{New Journal of Physics}}
  (\bibinfo{year}{2020}).

\bibitem{Ghatak2019Realization}
\bibinfo{author}{Ghatak, A.}, \bibinfo{author}{Brandenbourger, M.},
  \bibinfo{author}{van Wezel, J.} \& \bibinfo{author}{Coulais, C.}
\newblock \bibinfo{title}{Observation of non-hermitian topology and its
  bulk-edge correspondence} (\bibinfo{year}{2019}).
\newblock \eprint{arXiv:1907.11619v1}.

\bibitem{Zhou2020}
\bibinfo{author}{Zhou, D.} \& \bibinfo{author}{Zhang, J.}
\newblock \bibinfo{title}{Non-hermitian topological metamaterials with odd
  elasticity}.
\newblock \emph{\bibinfo{journal}{Phys. Rev. Research}}
  \textbf{\bibinfo{volume}{2}}, \bibinfo{pages}{023173} (\bibinfo{year}{2020}).

\bibitem{Scheibner2020}
\bibinfo{author}{Scheibner, C.} \emph{et~al.}
\newblock \bibinfo{title}{Odd elasticity}.
\newblock \emph{\bibinfo{journal}{Nature Physics}}
  \textbf{\bibinfo{volume}{16}}, \bibinfo{pages}{475--480}
  (\bibinfo{year}{2020}).

\bibitem{LiAlu2019}
\bibinfo{author}{Li, M.}, \bibinfo{author}{Ni, X.}, \bibinfo{author}{Weiner,
  M.}, \bibinfo{author}{Al\`u, A.} \& \bibinfo{author}{Khanikaev, A.~B.}
\newblock \bibinfo{title}{Topological phases and nonreciprocal edge states in
  non-hermitian floquet insulators}.
\newblock \emph{\bibinfo{journal}{Phys. Rev. B}}
  \textbf{\bibinfo{volume}{100}}, \bibinfo{pages}{045423}
  (\bibinfo{year}{2019}).

\bibitem{Sounas2017}
\bibinfo{author}{Sounas, D.~L.} \& \bibinfo{author}{Alu, A.}
\newblock \bibinfo{title}{Non-reciprocal photonics based on time modulation}.
\newblock \emph{\bibinfo{journal}{Nature Photonics}}
  \textbf{\bibinfo{volume}{11}}, \bibinfo{pages}{774} (\bibinfo{year}{2017}).

\bibitem{kotwal2019}
\bibinfo{author}{Kotwal, T.} \emph{et~al.}
\newblock \bibinfo{title}{Active topolectrical circuits}
  (\bibinfo{year}{2019}).
\newblock \eprint{1903.10130}.

\bibitem{helbig2019}
\bibinfo{author}{Helbig, T.} \emph{et~al.}
\newblock \bibinfo{title}{Observation of bulk boundary correspondence breakdown
  in topolectrical circuits} (\bibinfo{year}{2019}).
\newblock \eprint{1907.11562}.

\bibitem{Hofmann2020}
\bibinfo{author}{Hofmann, T.} \emph{et~al.}
\newblock \bibinfo{title}{Reciprocal skin effect and its realization in a
  topolectrical circuit}.
\newblock \emph{\bibinfo{journal}{Phys. Rev. Research}}
  \textbf{\bibinfo{volume}{2}}, \bibinfo{pages}{023265} (\bibinfo{year}{2020}).

\bibitem{Yoshida2020el}
\bibinfo{author}{Yoshida, T.}, \bibinfo{author}{Mizoguchi, T.} \&
  \bibinfo{author}{Hatsugai, Y.}
\newblock \bibinfo{title}{Mirror skin effect and its electric circuit
  simulation}.
\newblock \emph{\bibinfo{journal}{Phys. Rev. Research}}
  \textbf{\bibinfo{volume}{2}}, \bibinfo{pages}{022062} (\bibinfo{year}{2020}).

\bibitem{Wang2019}
\bibinfo{author}{Wang, Y.-X.} \& \bibinfo{author}{Clerk, A.~A.}
\newblock \bibinfo{title}{Non-hermitian dynamics without dissipation in quantum
  systems}.
\newblock \emph{\bibinfo{journal}{Phys. Rev. A}} \textbf{\bibinfo{volume}{99}},
  \bibinfo{pages}{063834} (\bibinfo{year}{2019}).

\bibitem{Lau2018}
\bibinfo{author}{Lau, H.-K.} \& \bibinfo{author}{Clerk, A.~A.}
\newblock \bibinfo{title}{Fundamental limits and non-reciprocal approaches in
  non-hermitian quantum sensing}.
\newblock \emph{\bibinfo{journal}{Nature Communications}}
  \textbf{\bibinfo{volume}{9}}, \bibinfo{pages}{4320} (\bibinfo{year}{2018}).

\bibitem{Bergholtz2019}
\bibinfo{author}{J., B.~E.}, \bibinfo{author}{Budich, J.~C.} \&
  \bibinfo{author}{Kunst, F.~K.}
\newblock \bibinfo{title}{Exceptional topology of non-hermitian systems}
  (\bibinfo{year}{2019}).
\newblock \eprint{arXiv:1912.10048}.

\bibitem{ashida2020}
\bibinfo{author}{Ashida, Y.}, \bibinfo{author}{Gong, Z.} \&
  \bibinfo{author}{Ueda, M.}
\newblock \bibinfo{title}{Non-hermitian physics} (\bibinfo{year}{2020}).
\newblock \eprint{2006.01837}.

\bibitem{Hatano1997}
\bibinfo{author}{Hatano, N.} \& \bibinfo{author}{Nelson, D.~R.}
\newblock \bibinfo{title}{Localization transitions in non-hermitian quantum
  mechanics}.
\newblock \emph{\bibinfo{journal}{Phys. Rev. Lett.}}
  \textbf{\bibinfo{volume}{77}}, \bibinfo{pages}{570--573}
  (\bibinfo{year}{1996}).

\bibitem{YaoSong2018}
\bibinfo{author}{Yao, S.}, \bibinfo{author}{Song, F.} \& \bibinfo{author}{Wang,
  Z.}
\newblock \bibinfo{title}{Non-hermitian chern bands}.
\newblock \emph{\bibinfo{journal}{Phys. Rev. Lett.}}
  \textbf{\bibinfo{volume}{121}}, \bibinfo{pages}{136802}
  (\bibinfo{year}{2018}).

\bibitem{Yao2018}
\bibinfo{author}{Yao, S.} \& \bibinfo{author}{Wang, Z.}
\newblock \bibinfo{title}{Edge states and topological invariants of
  non-hermitian systems}.
\newblock \emph{\bibinfo{journal}{Phys. Rev. Lett.}}
  \textbf{\bibinfo{volume}{121}}, \bibinfo{pages}{086803}
  (\bibinfo{year}{2018}).

\bibitem{Kunst2019}
\bibinfo{author}{Kunst, F.~K.} \& \bibinfo{author}{Dwivedi, V.}
\newblock \bibinfo{title}{Non-hermitian systems and topology: A transfer-matrix
  perspective}.
\newblock \emph{\bibinfo{journal}{Phys. Rev. B}} \textbf{\bibinfo{volume}{99}},
  \bibinfo{pages}{245116} (\bibinfo{year}{2019}).

\bibitem{Kunst2018}
\bibinfo{author}{Kunst, F.~K.}, \bibinfo{author}{Edvardsson, E.},
  \bibinfo{author}{Budich, J.~C.} \& \bibinfo{author}{Bergholtz, E.~J.}
\newblock \bibinfo{title}{Biorthogonal bulk-boundary correspondence in
  non-hermitian systems}.
\newblock \emph{\bibinfo{journal}{Phys. Rev. Lett.}}
  \textbf{\bibinfo{volume}{121}}, \bibinfo{pages}{026808}
  (\bibinfo{year}{2018}).

\bibitem{Torres2019}
\bibinfo{author}{Torres, L. E. F.~F.}
\newblock \bibinfo{title}{Perspective on topological states of non-hermitian
  lattices}.
\newblock \emph{\bibinfo{journal}{Journal of Physics: Materials}}
  \textbf{\bibinfo{volume}{3}}, \bibinfo{pages}{014002} (\bibinfo{year}{2019}).

\bibitem{Shen2018}
\bibinfo{author}{Shen, H.}, \bibinfo{author}{Zhen, B.} \& \bibinfo{author}{Fu,
  L.}
\newblock \bibinfo{title}{Topological band theory for non-hermitian
  hamiltonians}.
\newblock \emph{\bibinfo{journal}{Phys. Rev. Lett.}}
  \textbf{\bibinfo{volume}{120}}, \bibinfo{pages}{146402}
  (\bibinfo{year}{2018}).

\bibitem{Li2019}
\bibinfo{author}{Li, L.}, \bibinfo{author}{Lee, C.~H.} \&
  \bibinfo{author}{Gong, J.}
\newblock \bibinfo{title}{Geometric characterization of non-hermitian
  topological systems through the singularity ring in pseudospin vector space}.
\newblock \emph{\bibinfo{journal}{Phys. Rev. B}}
  \textbf{\bibinfo{volume}{100}}, \bibinfo{pages}{075403}
  (\bibinfo{year}{2019}).

\bibitem{Herviou2019}
\bibinfo{author}{Herviou, L.}, \bibinfo{author}{Bardarson, J.~H.} \&
  \bibinfo{author}{Regnault, N.}
\newblock \bibinfo{title}{Defining a bulk-edge correspondence for non-hermitian
  hamiltonians via singular-value decomposition}.
\newblock \emph{\bibinfo{journal}{Phys. Rev. A}} \textbf{\bibinfo{volume}{99}},
  \bibinfo{pages}{052118} (\bibinfo{year}{2019}).

\bibitem{Kawabata2019PRX}
\bibinfo{author}{Kawabata, K.}, \bibinfo{author}{Shiozaki, K.},
  \bibinfo{author}{Ueda, M.} \& \bibinfo{author}{Sato, M.}
\newblock \bibinfo{title}{Symmetry and topology in non-hermitian physics}.
\newblock \emph{\bibinfo{journal}{Phys. Rev. X}} \textbf{\bibinfo{volume}{9}},
  \bibinfo{pages}{041015} (\bibinfo{year}{2019}).

\bibitem{Kawabata2019NatCom}
\bibinfo{author}{Kawabata, K.}, \bibinfo{author}{Higashikawa, S.},
  \bibinfo{author}{Gong, Z.}, \bibinfo{author}{Ashida, Y.} \&
  \bibinfo{author}{Ueda, M.}
\newblock \bibinfo{title}{Topological unification of time-reversal and
  particle-hole symmetries in non-hermitian physics}.
\newblock \emph{\bibinfo{journal}{Nature Communications}}
  \textbf{\bibinfo{volume}{10}}, \bibinfo{pages}{297} (\bibinfo{year}{2019}).

\bibitem{Gong2018}
\bibinfo{author}{Gong, Z.} \emph{et~al.}
\newblock \bibinfo{title}{Topological phases of non-hermitian systems}.
\newblock \emph{\bibinfo{journal}{Phys. Rev. X}} \textbf{\bibinfo{volume}{8}},
  \bibinfo{pages}{031079} (\bibinfo{year}{2018}).

\bibitem{Ghatak2019}
\bibinfo{author}{Ghatak, A.} \& \bibinfo{author}{Das, T.}
\newblock \bibinfo{title}{New topological invariants in non-hermitian systems}.
\newblock \emph{\bibinfo{journal}{Journal of Physics: Condensed Matter}}
  \textbf{\bibinfo{volume}{31}}, \bibinfo{pages}{263001}
  (\bibinfo{year}{2019}).

\bibitem{Budich2019}
\bibinfo{author}{Budich, J.~C.}, \bibinfo{author}{Carlstr\"om, J.},
  \bibinfo{author}{Kunst, F.~K.} \& \bibinfo{author}{Bergholtz, E.~J.}
\newblock \bibinfo{title}{Symmetry-protected nodal phases in non-hermitian
  systems}.
\newblock \emph{\bibinfo{journal}{Phys. Rev. B}} \textbf{\bibinfo{volume}{99}},
  \bibinfo{pages}{041406} (\bibinfo{year}{2019}).

\bibitem{Jin2019Topological}
\bibinfo{author}{Zhang, K.~L.}, \bibinfo{author}{Wu, H.~C.},
  \bibinfo{author}{Jin, L.} \& \bibinfo{author}{Song, Z.}
\newblock \bibinfo{title}{Topological phase transition independent of system
  non-hermiticity}.
\newblock \emph{\bibinfo{journal}{Phys. Rev. B}}
  \textbf{\bibinfo{volume}{100}}, \bibinfo{pages}{045141}
  (\bibinfo{year}{2019}).

\bibitem{Lee2019}
\bibinfo{author}{Lee, C.~H.} \& \bibinfo{author}{Thomale, R.}
\newblock \bibinfo{title}{Anatomy of skin modes and topology in non-hermitian
  systems}.
\newblock \emph{\bibinfo{journal}{Phys. Rev. B}} \textbf{\bibinfo{volume}{99}},
  \bibinfo{pages}{201103} (\bibinfo{year}{2019}).

\bibitem{Lee2019b}
\bibinfo{author}{Lee, C.~H.}, \bibinfo{author}{li, L.},
  \bibinfo{author}{Thomale, R.} \& \bibinfo{author}{Gong, J.}
\newblock \bibinfo{title}{Unraveling non-hermitian pumping: emergent spectral
  singularities and anomalous responses} (\bibinfo{year}{2019}).
\newblock \eprint{arXiv:1912.06974}.

\bibitem{Borgnia2020}
\bibinfo{author}{Borgnia, D.~S.}, \bibinfo{author}{Kruchkov, A.~J.} \&
  \bibinfo{author}{Slager, R.-J.}
\newblock \bibinfo{title}{Non-hermitian boundary modes and topology}.
\newblock \emph{\bibinfo{journal}{Phys. Rev. Lett.}}
  \textbf{\bibinfo{volume}{124}}, \bibinfo{pages}{056802}
  (\bibinfo{year}{2020}).

\bibitem{Landau7}
\bibinfo{author}{Landau, L.} \emph{et~al.}
\newblock \emph{\bibinfo{title}{Theory of Elasticity}}.
\newblock Course of theoretical physics (\bibinfo{publisher}{Elsevier Science},
  \bibinfo{year}{1986}).

\bibitem{Paulose2015}
\bibinfo{author}{Paulose, J.}, \bibinfo{author}{Chen, B. G.-g.} \&
  \bibinfo{author}{Vitelli, V.}
\newblock \bibinfo{title}{{Topological modes bound to dislocations in
  mechanical metamaterials}}.
\newblock \emph{\bibinfo{journal}{Nat Phys}} \textbf{\bibinfo{volume}{11}},
  \bibinfo{pages}{153–--156} (\bibinfo{year}{2015}).

\bibitem{Paulose2015a}
\bibinfo{author}{Paulose, J.}, \bibinfo{author}{Meeussen, A.~S.} \&
  \bibinfo{author}{Vitelli, V.}
\newblock \bibinfo{title}{{Selective buckling via states of self-stress in
  topological metamaterials}}.
\newblock \emph{\bibinfo{journal}{Proceedings of the National Academy of
  Sciences}} \textbf{\bibinfo{volume}{112}}, \bibinfo{pages}{7639--7644}
  (\bibinfo{year}{2015}).

\bibitem{Rocklin2017}
\bibinfo{author}{Rocklin, D.~Z.}, \bibinfo{author}{Zhou, S.},
  \bibinfo{author}{Sun, K.} \& \bibinfo{author}{Mao, X.}
\newblock \bibinfo{title}{Transformable topological mechanical metamaterials}.
\newblock \emph{\bibinfo{journal}{Nature communications}}
  \textbf{\bibinfo{volume}{8}}, \bibinfo{pages}{14201} (\bibinfo{year}{2017}).

\bibitem{Chen2014}
\bibinfo{author}{Chen, B. G.-g.}, \bibinfo{author}{Upadhyaya, N.} \&
  \bibinfo{author}{Vitelli, V.}
\newblock \bibinfo{title}{{Nonlinear conduction via solitons in a topological
  mechanical insulator.}}
\newblock \emph{\bibinfo{journal}{Proceedings of the National Academy of
  Sciences of the United States of America}} \textbf{\bibinfo{volume}{111}},
  \bibinfo{pages}{13004--13009} (\bibinfo{year}{2014}).

\bibitem{Guest2003}
\bibinfo{author}{Guest, S.} \& \bibinfo{author}{Hutchinson, J.}
\newblock \bibinfo{title}{On the determinacy of repetitive structures}.
\newblock \emph{\bibinfo{journal}{Journal of the Mechanics and Physics of
  Solids}} \textbf{\bibinfo{volume}{51}}, \bibinfo{pages}{383 -- 391}
  (\bibinfo{year}{2003}).

\bibitem{Kane2014}
\bibinfo{author}{Kane, C.~L.} \& \bibinfo{author}{Lubensky, T.~C.}
\newblock \bibinfo{title}{{Topological boundary modes in isostatic lattices}}.
\newblock \emph{\bibinfo{journal}{Nature Physics}}
  \textbf{\bibinfo{volume}{10}}, \bibinfo{pages}{39--45}
  (\bibinfo{year}{2014}).

\bibitem{Fruchart2020}
\bibinfo{author}{Fruchart, M.}, \bibinfo{author}{Zhou, Y.} \&
  \bibinfo{author}{Vitelli, V.}
\newblock \bibinfo{title}{Dualities and non-abelian mechanics}.
\newblock \emph{\bibinfo{journal}{Nature}} \textbf{\bibinfo{volume}{577}},
  \bibinfo{pages}{636--640} (\bibinfo{year}{2020}).

\bibitem{Wiegmann2014}
\bibinfo{author}{Wiegmann, P.} \& \bibinfo{author}{Abanov, A.~G.}
\newblock \bibinfo{title}{Anomalous hydrodynamics of two-dimensional vortex
  fluids}.
\newblock \emph{\bibinfo{journal}{Phys. Rev. Lett.}}
  \textbf{\bibinfo{volume}{113}}, \bibinfo{pages}{034501}
  (\bibinfo{year}{2014}).

\bibitem{vanZuiden2016}
\bibinfo{author}{van Zuiden, B.~C.}, \bibinfo{author}{Paulose, J.},
  \bibinfo{author}{Irvine, W. T.~M.}, \bibinfo{author}{Bartolo, D.} \&
  \bibinfo{author}{Vitelli, V.}
\newblock \bibinfo{title}{Spatiotemporal order and emergent edge currents in
  active spinner materials}.
\newblock \emph{\bibinfo{journal}{Proc. Natl. Acad. Sci. USA}}
  \textbf{\bibinfo{volume}{113}}, \bibinfo{pages}{12919--12924}
  (\bibinfo{year}{2016}).

\bibitem{Han2020}
\bibinfo{author}{Han, M.} \emph{et~al.}
\newblock \bibinfo{title}{Statistical mechanics of a chiral active fluid}
  (\bibinfo{year}{2020}).
\newblock \eprint{2002.07679}.

\bibitem{Banerjee2017}
\bibinfo{author}{Banerjee, D.}, \bibinfo{author}{Souslov, A.},
  \bibinfo{author}{Abanov, A.~G.} \& \bibinfo{author}{Vitelli, V.}
\newblock \bibinfo{title}{Odd viscosity in chiral active fluids}.
\newblock \emph{\bibinfo{journal}{Nature Communications}}
  \textbf{\bibinfo{volume}{8}}, \bibinfo{pages}{1573} (\bibinfo{year}{2017}).

\bibitem{Galda2016}
\bibinfo{author}{Galda, A.} \& \bibinfo{author}{Vinokur, V.~M.}
\newblock \bibinfo{title}{Parity-time symmetry breaking in magnetic systems}.
\newblock \emph{\bibinfo{journal}{Phys. Rev. B}} \textbf{\bibinfo{volume}{94}},
  \bibinfo{pages}{020408} (\bibinfo{year}{2016}).

\bibitem{Nash2015}
\bibinfo{author}{Nash, L.~M.} \emph{et~al.}
\newblock \bibinfo{title}{{Topological mechanics of gyroscopic metamaterials.}}
\newblock \emph{\bibinfo{journal}{Proc. Natl. Acad. Sci. USA}}
  \textbf{\bibinfo{volume}{112}}, \bibinfo{pages}{14495--500}
  (\bibinfo{year}{2015}).

\bibitem{Wang2015}
\bibinfo{author}{Wang, P.}, \bibinfo{author}{Lu, L.} \&
  \bibinfo{author}{Bertoldi, K.}
\newblock \bibinfo{title}{{Topological Phononic Crystals with One-Way Elastic
  Edge Waves.}}
\newblock \emph{\bibinfo{journal}{Physical review letters}}
  \textbf{\bibinfo{volume}{115}}, \bibinfo{pages}{104302}
  (\bibinfo{year}{2015}).

\bibitem{Mostafazadeh2015}
\bibinfo{author}{Mostafazadeh, A.}
\newblock \bibinfo{title}{Physics of spectral singularities}.
\newblock In \bibinfo{editor}{Kielanowski, P.}, \bibinfo{editor}{Bieliavsky,
  P.}, \bibinfo{editor}{Odzijewicz, A.}, \bibinfo{editor}{Schlichenmaier, M.}
  \& \bibinfo{editor}{Voronov, T.} (eds.) \emph{\bibinfo{booktitle}{Geometric
  Methods in Physics}}, \bibinfo{pages}{145--165} (\bibinfo{publisher}{Springer
  International Publishing}, \bibinfo{address}{Cham}, \bibinfo{year}{2015}).

\bibitem{Bender1998}
\bibinfo{author}{Bender, C.~M.} \& \bibinfo{author}{Boettcher, S.}
\newblock \bibinfo{title}{Real spectra in non-hermitian hamiltonians having
  $pt$ symmetry}.
\newblock \emph{\bibinfo{journal}{Phys. Rev. Lett.}}
  \textbf{\bibinfo{volume}{80}}, \bibinfo{pages}{5243--5246}
  (\bibinfo{year}{1998}).

\bibitem{Fruchart2020phase}
\bibinfo{author}{Fruchart, M.}, \bibinfo{author}{Hanai, R.},
  \bibinfo{author}{Littlewood, P.~B.} \& \bibinfo{author}{Vitelli, V.}
\newblock \bibinfo{title}{Phase transitions in non-reciprocal active systems}
  (\bibinfo{year}{2020}).
\newblock \eprint{2003.13176}.

\bibitem{Haldane1988}
\bibinfo{author}{Haldane, F. D.~M.}
\newblock \bibinfo{title}{{Model for a Quantum Hall Effect without Landau
  Levels: Condensed-Matter Realization of the ``Parity Anomaly"}}.
\newblock \emph{\bibinfo{journal}{Physical Review Letters}}
  \textbf{\bibinfo{volume}{61}}, \bibinfo{pages}{2015--2018}
  (\bibinfo{year}{1988}).

\bibitem{Kawabata2019PRL}
\bibinfo{author}{Kawabata, K.}, \bibinfo{author}{Bessho, T.} \&
  \bibinfo{author}{Sato, M.}
\newblock \bibinfo{title}{Classification of exceptional points and
  non-hermitian topological semimetals}.
\newblock \emph{\bibinfo{journal}{Phys. Rev. Lett.}}
  \textbf{\bibinfo{volume}{123}}, \bibinfo{pages}{066405}
  (\bibinfo{year}{2019}).

\bibitem{Zhou2019}
\bibinfo{author}{Zhou, H.}, \bibinfo{author}{Lee, J.~Y.}, \bibinfo{author}{Liu,
  S.} \& \bibinfo{author}{Zhen, B.}
\newblock \bibinfo{title}{Exceptional surfaces in pt-symmetric non-hermitian
  photonic systems}.
\newblock \emph{\bibinfo{journal}{Optica}} \textbf{\bibinfo{volume}{6}},
  \bibinfo{pages}{190--193} (\bibinfo{year}{2019}).

\bibitem{Xiong2018}
\bibinfo{author}{Xiong, Y.}
\newblock \bibinfo{title}{Why does bulk boundary correspondence fail in some
  non-hermitian topological models}.
\newblock \emph{\bibinfo{journal}{Journal of Physics Communications}}
  \textbf{\bibinfo{volume}{2}}, \bibinfo{pages}{035043} (\bibinfo{year}{2018}).

\bibitem{Heiss2012}
\bibinfo{author}{Heiss, W.}
\newblock \bibinfo{title}{The physics of exceptional points}.
\newblock \emph{\bibinfo{journal}{Journal of Physics A: Mathematical and
  Theoretical}} \textbf{\bibinfo{volume}{45}}, \bibinfo{pages}{444016}
  (\bibinfo{year}{2012}).

\bibitem{Yoshida2019qu}
\bibinfo{author}{Yoshida, T.}, \bibinfo{author}{Peters, R.},
  \bibinfo{author}{Kawakami, N.} \& \bibinfo{author}{Hatsugai, Y.}
\newblock \bibinfo{title}{Symmetry-protected exceptional rings in
  two-dimensional correlated systems with chiral symmetry}.
\newblock \emph{\bibinfo{journal}{Phys. Rev. B}} \textbf{\bibinfo{volume}{99}},
  \bibinfo{pages}{121101} (\bibinfo{year}{2019}).

\bibitem{Okugawa2019}
\bibinfo{author}{Okugawa, R.} \& \bibinfo{author}{Yokoyama, T.}
\newblock \bibinfo{title}{Topological exceptional surfaces in non-hermitian
  systems with parity-time and parity-particle-hole symmetries}.
\newblock \emph{\bibinfo{journal}{Phys. Rev. B}} \textbf{\bibinfo{volume}{99}},
  \bibinfo{pages}{041202} (\bibinfo{year}{2019}).

\bibitem{Vitelli2010}
\bibinfo{author}{Vitelli, V.}, \bibinfo{author}{Xu, N.},
  \bibinfo{author}{Wyart, M.}, \bibinfo{author}{Liu, A.~J.} \&
  \bibinfo{author}{Nagel, S.~R.}
\newblock \bibinfo{title}{Heat transport in model jammed solids}.
\newblock \emph{\bibinfo{journal}{Phys. Rev. E}} \textbf{\bibinfo{volume}{81}},
  \bibinfo{pages}{021301} (\bibinfo{year}{2010}).

\bibitem{benzoni2020}
\bibinfo{author}{Benzoni, C.}, \bibinfo{author}{Jeevanesan, B.} \&
  \bibinfo{author}{Moroz, S.}
\newblock \bibinfo{title}{Rayleigh edge waves in two-dimensional chiral
  crystals} (\bibinfo{year}{2020}).
\newblock \eprint{2004.09517}.

\bibitem{Beatus2006}
\bibinfo{author}{Beatus, T.}, \bibinfo{author}{Tlusty, T.} \&
  \bibinfo{author}{Bar-Ziv, R.}
\newblock \bibinfo{title}{Phonons in a one-dimensional microfluidic crystal}.
\newblock \emph{\bibinfo{journal}{Nature Physics}}
  \textbf{\bibinfo{volume}{2}}, \bibinfo{pages}{743--748}
  (\bibinfo{year}{2006}).

\bibitem{Prost2015}
\bibinfo{author}{Prost, J.}, \bibinfo{author}{J{\"u}licher, F.} \&
  \bibinfo{author}{Joanny, J.}
\newblock \bibinfo{title}{Active gel physics}.
\newblock \emph{\bibinfo{journal}{Nature Physics}}
  \textbf{\bibinfo{volume}{11}}, \bibinfo{pages}{111} (\bibinfo{year}{2015}).

\bibitem{Souslov2017}
\bibinfo{author}{Souslov, A.}, \bibinfo{author}{van Zuiden, B.~C.},
  \bibinfo{author}{Bartolo, D.} \& \bibinfo{author}{Vitelli, V.}
\newblock \bibinfo{title}{Topological sound in active-liquid metamaterials}.
\newblock \emph{\bibinfo{journal}{Nature Physics}}
  \textbf{\bibinfo{volume}{13}}, \bibinfo{pages}{1091} (\bibinfo{year}{2017}).

\bibitem{White1962}
\bibinfo{author}{White, D.~L.}
\newblock \bibinfo{title}{Amplification of ultrasonic waves in piezoelectric
  semiconductors}.
\newblock \emph{\bibinfo{journal}{Journal of Applied Physics}}
  \textbf{\bibinfo{volume}{33}}, \bibinfo{pages}{2547--2554}
  (\bibinfo{year}{1962}).

\bibitem{Lakes2009}
\bibinfo{author}{Lakes, R.}
\newblock \emph{\bibinfo{title}{Viscoelastic Materials}}
  (\bibinfo{publisher}{Cambridge University Press}, \bibinfo{year}{2009}).

\bibitem{Souslov2020PRE}
\bibinfo{author}{Souslov, A.}, \bibinfo{author}{Gromov, A.} \&
  \bibinfo{author}{Vitelli, V.}
\newblock \bibinfo{title}{Anisotropic odd viscosity via a time-modulated
  drive}.
\newblock \emph{\bibinfo{journal}{Phys. Rev. E}}
  \textbf{\bibinfo{volume}{101}}, \bibinfo{pages}{052606}
  (\bibinfo{year}{2020}).

\bibitem{Soni2018}
\bibinfo{author}{Soni, V.} \emph{et~al.}
\newblock \bibinfo{title}{The free surface of a colloidal chiral fluid: waves
  and instabilities from odd stress and hall viscosity} (\bibinfo{year}{2018}).
\newblock \eprint{arXiv:1812.09990v1}.

\bibitem{Avron1998}
\bibinfo{author}{Avron, J.~E.}
\newblock \bibinfo{title}{{Odd Viscosity}}.
\newblock \emph{\bibinfo{journal}{Journal of Statistical Physics}}
  \textbf{\bibinfo{volume}{92}}, \bibinfo{pages}{543--557}
  (\bibinfo{year}{1998}).

\bibitem{Sone2019}
\bibinfo{author}{Sone, K.} \& \bibinfo{author}{Ashida, Y.}
\newblock \bibinfo{title}{Anomalous topological active matter}.
\newblock \emph{\bibinfo{journal}{Phys. Rev. Lett.}}
  \textbf{\bibinfo{volume}{123}}, \bibinfo{pages}{205502}
  (\bibinfo{year}{2019}).

\bibitem{Sone2019arxiv}
\bibinfo{author}{Sone, K.}, \bibinfo{author}{Ashida, Y.} \&
  \bibinfo{author}{Sagawa, T.}
\newblock \bibinfo{title}{Exceptional non-hermitian topological edge mode and
  its application to active matter}  (\bibinfo{year}{2019}).
\newblock \eprint{arXiv:1912.09055}.

\bibitem{Lakes2001}
\bibinfo{author}{Lakes, R.}
\newblock \bibinfo{title}{Elastic and viscoelastic behavior of chiral
  materials}.
\newblock \emph{\bibinfo{journal}{International Journal of Mechanical
  Sciences}} \textbf{\bibinfo{volume}{43}}, \bibinfo{pages}{1579 -- 1589}
  (\bibinfo{year}{2001}).

\bibitem{Lakes2016}
\bibinfo{author}{Lakes, R.~S.}
\newblock \bibinfo{title}{Physical meaning of elastic constants in cosserat,
  void, and microstretch elasticity}.
\newblock \emph{\bibinfo{journal}{Journal of Mechanics of Materials and
  Structures}} \textbf{\bibinfo{volume}{11}}, \bibinfo{pages}{217--229}
  (\bibinfo{year}{2016}).

\bibitem{Mitchell2018}
\bibinfo{author}{Mitchell, N.~P.}, \bibinfo{author}{Nash, L.~M.} \&
  \bibinfo{author}{Irvine, W. T.~M.}
\newblock \bibinfo{title}{Tunable band topology in gyroscopic lattices}.
\newblock \emph{\bibinfo{journal}{Phys. Rev. B}} \textbf{\bibinfo{volume}{98}},
  \bibinfo{pages}{174301} (\bibinfo{year}{2018}).

\bibitem{Mitchell2018Realization}
\bibinfo{author}{Mitchell, N.~P.}, \bibinfo{author}{Nash, L.~M.} \&
  \bibinfo{author}{Irvine, W. T.~M.}
\newblock \bibinfo{title}{Realization of a topological phase transition in a
  gyroscopic lattice}.
\newblock \emph{\bibinfo{journal}{Phys. Rev. B}} \textbf{\bibinfo{volume}{97}},
  \bibinfo{pages}{100302} (\bibinfo{year}{2018}).

\bibitem{Mitchell2018Nature}
\bibinfo{author}{Mitchell, N.~P.}, \bibinfo{author}{Nash, L.~M.},
  \bibinfo{author}{Hexner, D.}, \bibinfo{author}{Turner, A.~M.} \&
  \bibinfo{author}{Irvine, W. T.~M.}
\newblock \bibinfo{title}{Amorphous topological insulators constructed from
  random point sets}.
\newblock \emph{\bibinfo{journal}{Nature Physics}}
  \textbf{\bibinfo{volume}{14}}, \bibinfo{pages}{380--385}
  (\bibinfo{year}{2018}).

\bibitem{Bernard2002}
\bibinfo{author}{Bernard, D.} \& \bibinfo{author}{LeClair, A.}
\newblock \emph{\bibinfo{title}{A Classification of Non-Hermitian Random
  Matrices}}, \bibinfo{pages}{207--214} (\bibinfo{publisher}{Springer
  Netherlands}, \bibinfo{address}{Dordrecht}, \bibinfo{year}{2002}).

\bibitem{FRUCHART2013}
\bibinfo{author}{Fruchart, M.} \& \bibinfo{author}{Carpentier, D.}
\newblock \bibinfo{title}{An introduction to topological insulators}.
\newblock \emph{\bibinfo{journal}{Comptes Rendus Physique}}
  \textbf{\bibinfo{volume}{14}}, \bibinfo{pages}{779 -- 815}
  (\bibinfo{year}{2013}).
\newblock \bibinfo{note}{Topological insulators / Isolants topologiques}.

\bibitem{Fukui2013}
\bibinfo{author}{Fukui, T.}, \bibinfo{author}{Hatsugai, Y.} \&
  \bibinfo{author}{Suzuki, H.}
\newblock \bibinfo{title}{{Chern Numbers in Discretized Brillouin Zone:
  Efficient Method of Computing (Spin) Hall Conductances}}.
\newblock \emph{\bibinfo{journal}{Journal of the Physical Society of Japan}}
  (\bibinfo{year}{2013}).

\bibitem{Chen2019}
\bibinfo{author}{Chen, W.} \& \bibinfo{author}{Schnyder, A.~P.}
\newblock \bibinfo{title}{Universality classes of topological phase transitions
  with higher-order band crossing}.
\newblock \emph{\bibinfo{journal}{New Journal of Physics}}
  \textbf{\bibinfo{volume}{21}}, \bibinfo{pages}{073003}
  (\bibinfo{year}{2019}).

\end{thebibliography}
\end{document}